\documentclass{aastex631}
\usepackage{lineno}
\received{January 23, 2024}
\revised{April 4, 2024}
\revised{April 11, 2024}
\accepted{April 11, 2024}

\begin{document}

\title{Boron Abundances in Early B Dwarfs
of the Galactic Open Cluster NGC 3293
\footnote{Based on observations made with the NASA/ESA Hubble Space Telescope, obtained at the Space Telescope Science Institute, which is operated by the Association of Universities for Research in Astronomy, Inc., under NASA contract NAS 5-26555. These observations are associated with proposal GO-14673.}}


\author[0000-0001-7617-5665]{Charles R. Proffitt}
\affiliation{Space Telescope Science Institute \\
3700 San Martin Drive \\
Baltimore, MD 21210, USA}

\author[0000-0003-2946-9390]{Harim Jin}
\affiliation{Argelander-Institut f\"ur Astronomie, Bonn University, Auf dem H\"ugel 71, 53121, Bonn, Germany}

\author[0000-0001-9205-2307]{Simone Daflon}
\affiliation{Observatório Nacional - MCTI (ON)\\ Rua Gal. José Cristino 77, São Cristóvão\\ 20921-400 Rio de Janeiro, Brazil)}

\author[0000-0003-3063-4867]{Daniel J. Lennon}
\affiliation{Instituto de Astrofísica de Canarias\\ 38 200, La Laguna, Tenerife, Spain} 
\affiliation{Dpto. Astrofísica, Universidad de La Laguna\\ 38 205, La Laguna, Tenerife, Spain}

\author[0000-0003-3026-0367]{Norbert Langer}
\affiliation{Argelander-Institut f\"ur Astronomie, Bonn University, Auf dem H\"ugel 71, 53121, Bonn, Germany}

\author[0000-0001-6476-0576]{Katia Cunha}
\affiliation{University of Arizona\\ Tucson, AZ 85719, USA}
\affiliation{Observatório Nacional, São Cristóvão\\ Rio de Janeiro, Brazil}

\author[0000-0001-6797-6081]{Talawanda Monroe}
\affiliation{Space Telescope Science Institute \\
3700 San Martin Drive \\
Baltimore, MD 21210, USA}

\submitjournal{the Astrophysical Journal}

\begin{abstract}

New boron abundances or upper limits have been determined for 8 early-B stars in the young Galactic open cluster NGC 3293, using ultraviolet spectra obtained by the Hubble Space Telescope Cosmic Origins Spectrograph.  With previous observations, there are now 18 early-B stars in this cluster with boron measurements. Six of the newly observed stars have projected rotational velocities greater than 200 km/s, allowing new constraints on rotationally driven mixing in main-sequence stars. When comparing to synthetic model populations, we find that the majority of our sample stars agree well with the predicted trends of stronger boron depletion for larger rotation and for larger mass or luminosity. Based on those, a smaller than the canonical rotational mixing efficiency, ($f_c \approx 0.0165$ vs the more standard value of 0.033), appears to be { favored}.
In addition, the five mostly slowly rotating stars, { when considered as a group, tend to show more boron depletion than expected from rotational mixing, and we speculate that most or all of these} originate from binary mergers.

\end{abstract}


\section{Introduction} \label{sec:intro}

The dominant energy source in core hydrogen burning massive stars is the
fusion of protons to $^{4}$He. Since massive stars use the CNO-cycle to perform this, 
most of the available carbon and oxygen in the core is converted into nitrogen during this process.
At the same time, the pp-chains are also operating. While they contribute little to the energy production
in massive stars, the corresponding proton capture reactions leads to the destruction of the light elements, 
lithium, beryllium and boron, in most of the stellar interior.

It has been known for several decades, that the {\em surfaces} of a fraction of the upper massive main-sequence stars 
show abundance patterns which indicate enrichment with hydrogen burning products.
These include anti-correlated nitrogen and carbon abundances indicative of CN processing \citep{1988A&A...197..209S, 1992ApJ...387..673G, 2010A&A...517A..38P}, enhanced helium \citep{1992LNP...401...21H}, and boron depletion \citep{2001ApJ...548..429P, 2002ApJ...565..571V, 2006ApJ...640.1039M}. It has been suggested that these abundance changes have been caused by rotationally driven internal mixing \citep{2010NewAR..54...32M, 2012ARA&A..50..107L}, and many stellar evolution models for massive stars do now include its effects  \citep{2011A&A...530A.115B, 2012A&A...537A.146E, 2013ApJS..208....4P}.

However, refined observations cast doubts on this explanation. \citet{2008ApJ...676L..29H}, identified a population of slowly rotating early B\,dwarfs in the LMC which showed a significant surface nitrogen enrichment. A similar population was also found in the LMC O\,star population \citep{2012A&A...537A..79R, 2017A&A...600A..82G}.
{ Similar discrepancies have been found in Galactic stars. Some well studied stars that are known to be intrinsically slow rotators nevertheless show evidence of mixing, \citep[e.g., $\tau$\ Sco as discussed by][]{2021MNRAS.504.2474K}, while  \citet{2017A&A...604A.123C} finds fast rotators without the expected nitrogen enrichment, and \citet{2014A&A...562A..37M, 2018A&A...613A..12M}, discuss slowly rotating nitrogen rich stars.} 
At the same time, it became clear that a typical population of massive main sequence single stars must be expected to contain a significant fraction ($\sim 20$\%) of binary 
products, i.e., mass gainers in mass transfer systems or mergers, \citep{2012Sci...337..444S, 2014ApJ...782....7D}. 

In this paper, we pursue the idea that boron abundance measurements may have a strong potential
to discriminate between the suggested enrichment channels.
\citet{2016ApJ...824....3P} discussed observations of the \ion{B}{3}\ resonance line in 10 early-B stars in the young open cluster NGC~3293, and concluded that modest boron depletion was seen, { and that for most stars the depletion was slightly less efficient than predicted by} the models of \cite{2011A&A...530A.115B}, in which rotational mixing was calibrated using observations of nitrogen abundances in massive main sequence stars of the LMC.  { In particular they noted that three of the ten stars appeared to be slightly boron rich relative to the expectations of these models.} We requested observations of ten additional stars, and obtained useful data for 8 of these, as part of {\em HST} program 14673, with a focus on more rapidly rotating stars lower on the main-sequence. These data provide us with a unique sample of 18 early B dwarfs with information on their boron and nitrogen abundances in an open star cluster, i.e., with roughly known and equal ages.

\section{UV observations}

For this analysis, we will combine the new data obtained in {\em HST} GO program 14673, with the older data from program {\em HST} GO 12520. The latter observations were described in \citet{2016ApJ...824....3P}. The stars observed by both programs are listed in Table \ref{tab:uvobs} along with descriptive information for each star from \citet{2006Dufton}. The observational strategy for 14673 was very similar to that of 12520, utilizing the COS G185M 1971 setting.  This setting covers three distinct wavelength ranges, the third of which covers from 2052 to 2087 \AA\ which includes the \ion{B}{3}\ resonance doublet. The observations for 14673 were executed in September of 2016 and June of 2017. 

The reduction of the UV spectra also followed the procedures detailed in \citet{2016ApJ...824....3P}, including the need to correct for the variable vignetting of up to 20\% that affects the short wavelength end of each COS NUV "stripe", \citep{2010cos..rept....3A}. This needs to be done by comparing the data taken at each individual FP-POS position to derive a correction before combining them into a final merged spectrum.  This correction is not included in the standard pipeline products available from the MAST archive.

Targets for both programs were selected from \cite{2006Dufton}, based on their location in the HR diagram as well as their rotational velocity. For the new program, we gave preference to more rapidly rotating stars.
{ An additional important criterion was cluster membership, as implied by both photometric and spectroscopic surveys of the cluster \citep{2003A&A...402..549B,2006Dufton}. Subsequently, \citet{2022A&A...660A..11G} have analysed $Gaia$ data for this cluster and confirm cluster membership for all sources considered here, barring possibly ESL\,028, that they comment on as a possible 2$\sigma$ background star based on its parallax measurement. This star has a RUWE value of 1.50, which is is out-of-family with other cluster members that have values close to 0.9, and based on its overall $Gaia$ properties it is still designated a cluster member by \citet{Morel2022}.}

{ One might consider using radial velocity information to further refine cluster membership probability. However as noted by \cite{1958MNRAS.118..618F} there are a significant number of radial velocity variables in the cluster.  Those data suggest that much of this variability is associated with $\beta$ Cep variables, which are common in this cluster, (see Table 1 for known variables), given the temperature of its turn-off stars. Moreover a preliminary analysis of ESO archival data for selected $\beta$ Cep members confirms significant apparent velocity shifts occur on short timescales, comparable to their periods. However we agree with \citet{1958MNRAS.118..618F} that ESL\,008 is a potential SB1 system with a peak-to-peak velocity of $\sim$90\,km/s, cluster membership implying a minimum semi-amplitude velocity of $\sim$70\,km/s. Note that while \citet{Morel2022} suggest this source to be an SB2 system, their analysis is based on a single VLT/UVES spectrum. However comparing their data with the 2.2m/FEROS data of \citet{2006Dufton} we find no evidence of a secondary in either spectrogram.
}

After the UV observations were obtained, we found that ESL 016 was a much narrower lined star than we had expected. While \citet{2006Dufton} had adopted the correct photometric information for this $V=9.23$ star, the FLAMES-Giraffe Spectrograph observation that they associated with it appears to have actually been of the fainter, ($V=11.55$), star NGC 3293 B 198, (apparently Gaia EDR3 5351447238427079936), \citep{1994MNRAS.267.1060B}, which is located about 2" away from ESL 16. {\citet{Morel2022}, appears to have adopted the same mis-identification}. Fortunately, the ESO archive does contain three similar FEROS observations of the brighter star, under the name V 404 Car, taken in 2016 by F. Rodler as part of ESO program 079.A-9008, and we were able to use these spectra for our analysis.

We summarize the UV observations of each star in Table \ref{tab:uvobs}, along with basic information for each star from \citet{2006Dufton}. We include the signal-to-noise achieved at the location of the \ion{B}{3} line over a 0.1\AA\ bin, which corresponds to the $\approx 3$ pixel COS G185M resolution element.  We also note whether a star is a known variable, several are ${\beta}$ Cehpehi stars, and whether there is any evidence for radial velocity variations.  Most of our targets appear to currently be single stars.

\begin{deluxetable}{crlcccl}
\tablecaption{Stars observed in programs 12520 and 14673.\tablenotemark{a}\label{tab:uvobs}}
\tablehead{\colhead{ESL no.}  &  \colhead{$V$} &  \colhead{Sp.\ type}  & \colhead{$V \sin i$} & \colhead{HST PID} & S/N per &\colhead{comments} \\
                  &        &          & \colhead{(km$/$s)}&         & 0.1\AA & \\}
\startdata
02&  6.73 &  B0.7 Ib    &     100   &  12520& 221&   STIS E230H\\
03&  7.61 &  B1 III     &      80   &  12520& 304&  \\
04&  8.03 &  B1 III     &     105   &  12520& 320&  \\
05&  8.12 &  B1 III     &     195   &  14673& 246& \\
06&  8.21 &  B1 III     &     200   &  14673& 245& \\
07&  8.25 &  B1 III     &      65   &  12520& 185&  \\
08&  8.59 &  B1 III     &     140   &  12520& 220& RV var. from \citet{1958MNRAS.118..618F} \\
10&  8.77 &  B1 III     &      70   &  12520& 252& V403 Car, $P=0.2506$d \\
12&  8.95 &  B1 III     &     100   &  12520& 179& V380 Car, $P=0.2270$d \\
15&  9.11 &  B1 V       &     260   &  14673& 207& V440 Car, $P=0.1790$d \\
16&  9.21 &  BI III     &      10   &  14673& 172& V404 Car, $P=0.1621$d \\
19&  9.27 &  B1 V       &     120   &  12520& 223& V405 Car, $P=0.1524$d; SB2 \\
20&  9.55 &  B1.5 III   &      60   &  14673& 50& V401 Car, $P=0.1684$d; bad pointing no useful data \\
23& 10.01 &  B1.5 III   &     160   &  12520& 159&  \\
24& 10.01 &  B1.5 III   &     135   &  12520& 147& RV var from \citet{1958MNRAS.118..618F} \\
25& 10.01 &  B2 III     &     215   &  14673& 124& no good fit to UV spectrum\\
28& 10.26 &  B2 V       &     215   &  14673& 177& \\
30& 10.51 &  B2 V       &     205   &  14673& 173& RV var from \citet{Morel2022} \\
31& 10.66 &  B2 V       &     230   &  14673& 133& \\
38& 11.00 &  B2.5 V    &     235   &  14673& 125& \\
\enddata
\tablenotetext{a}{Stellar parameters shown here are as given by  \citet{2006Dufton}, except for ESL 16 which is from \cite{2005ApJS..158..193S}, as the spectra of Dufton for this star were actually of a fainter companion.}
\end{deluxetable}

For one of the targets, ESL 20, a bad guide star acquisition caused a mis-pointed observation with the target on the edge of the aperture and only a very faint spectrum visible.  Since the program was over 90\% complete, this visit was not eligible for a repeat.  In addition, for ESL 25, the observed UV spectrum appears to be a composite that does not correspond to the optical spectrum and could not be usefully fit.  While this star does have a companion about 1.6 magnitudes fainter at a distance of about 2.3", that star should have been well out of the 2.5" diameter COS aperture, and the 2D spectral image shows no evidence for multiple components. As a result, these two stars will be left out of our analysis.

\section{Determination of Stellar Parameters}

\subsection{Discussion of Literature Results}

An extensive spectroscopic analysis of B stars in NGC 3293 was included as part of the FLAMES survey \citep{2005A&A...437..467E}, and the resulting stellar parameter and abundance determinations were presented in \citet{2006Dufton} and \citet{Hunter2009}. This work primarily relied upon H and He line profiles (H$\gamma$, H$\delta$, and \ion{He}{1} 4026\AA) to determine $T_{\rm eff}$ and $\log g$, while for some stars where the He lines were not useful, parameters were estimated based on the spectral type.  For the most narrow lined stars, the Si ionization ratios were also used to constrain the effective temperature. 

{
A recent paper by \citet{Morel2022} reanalyzed a subset of these stars and determined stellar parameters from a global fit to to several spectral regions that included H$\gamma$, H$\delta$, as well as various lines of \ion{He}{1}, \ion{Si}{2}, \ion{Si}{3}, \ion{O}{2} that put useful constraints on the stellar parameters. They then determined abundances from selected lines of individual elements, including \ion{C}{2}\ $\lambda$4267 and \ion{N}{2}\ $\lambda$4630. \citet{Morel2022}, however, did not attempt to measure the value of the microturbulence, $\xi$, instead using fixed values of 2 or 5 km/s depending on stellar parameters.}


{
For the stars we consider, we show the results for stellar parameters, carbon, and nitrogen abundances found by both \citet{Hunter2009} and  \citet{Morel2022} in Table \ref{tab:prevwork}, and will discuss their findings in the context of our new results below. For those results presented by  \citet{Hunter2009}\ where the parameters were estimated from the spectral type, we mark the $T_{\rm eff}$ with a colon to indicate the greater uncertainty in the parameters.}

\begin{deluxetable}{clccccccccccc}
\tablecaption{Previous Stellar Parameter and Abundance Determinations\label{tab:prevwork}}
\tablehead{\colhead{ESL\#} &
\multicolumn{6}{c}{\cite{2006Dufton} and  \cite{Hunter2009}\tablenotemark{a}} &
\multicolumn{6}{c}{Morel et al.\ (2022)}\\
 & \multicolumn{6}{c}{\rule{210pt}{1pt}   } & \multicolumn{6}{c}{\rule{210pt}{1pt} }\\
\colhead{} &
\colhead{$T_{ef\!f}$} & \colhead{$\log\,g$} & \colhead{$\xi$} & \colhead{$v\sin i$}& \colhead{C}& \colhead{N} &
\colhead{$T_{ef\!f}$} & \colhead{$\log\,g$} & \colhead{$\xi$} & \colhead{$v\sin i$}& \colhead{C}& \colhead{N} 
} 
\startdata
002 & 23200& 2.80 &  15 &  100  & $8.10\pm 0.22$    & $8.08\pm 0.12$  &\nodata&\nodata&\nodata&\nodata&\nodata&\nodata\\
003 & 20500& 2.75 &  13 &   80  & $7.95\pm 0.22$    & $7.52\pm 0.06$  &\nodata&\nodata&\nodata&\nodata&\nodata&\nodata\\
004 & 22700& 3.13 &  13 &  105  & $8.17\pm 0.28$    & $7.55\pm 0.09$  &\nodata&\nodata&\nodata&\nodata&\nodata&\nodata\\
005 & 21500:& 3.05 &  14 &  195  & $7.89\pm 0.24$    & $7.56\pm 0.08$  &\nodata&\nodata&\nodata&\nodata&\nodata&\nodata\\
006 & 21500:& 3.15 &  15 &  200  & $7.81\pm 0.23$    & $7.37\pm 0.16$  &\nodata&\nodata&\nodata&\nodata&\nodata&\nodata\\
007 & 22600& 3.10 &  11 &   65  & $8.13\pm 0.22$    & $7.50\pm 0.08$  &\nodata&\nodata&\nodata&\nodata&\nodata&\nodata\\
008 & 21500:& 3.25 &  12 &  140  & $7.83\pm 0.23$    & $7.59\pm 0.15$  &\nodata&\nodata&\nodata&\nodata&\nodata&\nodata\\   
010 & 21450& 3.20 &  11 &   70  & $7.85\pm 0.21$    & $7.45\pm 0.08$  &    24493 &   3.61 &5  &   41 &    $8.19\pm 0.16$  &    $7.57\pm 0.14$   \\
012 & 21500:& 3.30 &  11 &  100  & $7.86\pm 0.17$    & $7.45\pm 0.13$  &    24444 &   3.70 &5  &   83 &    $8.41\pm 0.23$  &    $7.63\pm 0.14$   \\
015 & 25000:& 3.80 &\nodata&  260  &\nodata&\nodata&\nodata&\nodata&\nodata&\nodata&\nodata&\nodata\\
019 & 25000:& 3.85 &   6 &  120  & $8.26\pm 0.29$    & $7.59\pm 0.25$  &    24380 &   3.63 &5  &  107 &    $8.06\pm 0.23$  &    $7.63\pm 0.20$   \\
023 & 20500:& 3.40 &   6 &  160  & $7.72\pm 0.25$    & $7.57\pm 0.23$  &    21730 &   3.71 &2  &  154 &    $7.99\pm 0.15$  &    $7.94\pm 0.20$   \\
024 & 20500:& 3.50 &   6 &  135  & $7.75\pm 0.23$    & $7.62\pm 0.25$  &    21700 &   3.69 &2  &  132 &    $8.03\pm 0.23$  &    $7.84\pm 0.20$   \\
025 & 21100& 3.70 &   5 &  215  & $7.84\pm 0.43$    & $7.41\pm 0.48$  &    20460 &   3.69 &2  &  192 &    $8.17\pm 0.23$  &    $7.64\pm 0.30$   \\
028 & 19400& 3.65 &   7 &  215  & $7.69\pm 0.33$    & $7.52\pm 0.30$  &    20429 &   3.86 &2  &  165 &    $7.80\pm 0.23$  &    $7.62\pm 0.21$   \\
030 & 19800& 3.70 &   4 &  205  & $7.86\pm 0.27$    & $7.57\pm 0.27$  &    19675 &   3.80 &2  &  216 &    $8.24\pm 0.11$  &    $8.01\pm 0.12$   \\
031 & 17400& 3.45 &  12 &  230  & $7.66\pm 0.33$    & $7.70\pm 0.37$  &    20322 &   3.86 &2  &  257 &    $8.15\pm 0.11$  &    $7.66\pm 0.14$   \\
038 & 20600& 3.95 &   1 &  235  & $8.16\pm 0.26$    & $7.57\pm 0.29$  &    20216 &   3.94 &2  &  229 &    $8.20\pm 0.11$  &    $7.65\pm 0.13$ 
\enddata
\tablenotetext{a}{For most stars \citet{Hunter2009} adopted the stellar parameters from \citet{2006Dufton}, but slightly updated the parameters for a few stars in addition to deriving abundances. $T_{\rm eff}$ values which \citet{2006Dufton} derived completely or partially from estimated spectral types have larger uncertainties and are marked with a colon.}

\end{deluxetable}

\subsection{New Analysis of Optical Spectra\label{sect:optanal}}

To determine revised stellar parameters, we reanalyzed the same optical spectra used by \citet{2006Dufton}. These observations were originally described in \citet{2005A&A...437..467E}. For the brighter stars, (ESL 02 through ESL 19), in our sample, these spectra were obtained with the Fibre-Fed Extended Range Optical Spectrograph (FEROS) at the Max Planck Gesellschaft MPG/ESO telescope, with $R \approx 48000$ and coverage between 3600 and 9200 \AA), while for the remainder of the stars, the Fibre Large Array Multi-Element Spectrograph (FLAMES) instrument at the Very Large Telescope (VLT) was used with the FLAMES-Giraffe Spectrograph ($R \approx 25000$). { Giraffe settings HR02, (cenwave $\lambda$3958), HR03 ($\lambda$4124), HR04 ($\lambda$4297), HR05 ($\lambda$4471), HR06 ($\lambda$4656), and HR14 ($\lambda$6515) were used providing coverage from 3850--4475\AA\ and 6380--6620\AA\ were used.
}

{ 
In this study, we selected the four sample stars in NGC 3293, having the lowest $V \sin i$, as benchmark stars for which we could analyze a large number of spectral lines. The selected stars were: ESL 02, 03, 07, and 10. We then used the wrapper  Stellar Spectral Synthesis Suite (S4,   \citet{Bragancga2019}) to determine their $T_{\rm eff}$, $\log g$, the abundances of silicon and oxygen, as well as the projected rotational velocity $V \sin i$, the radial-tangential macroturbulent velocity $\zeta$, and the microturbulent velocity $\xi$. The iterative scheme of S4   interpolates in a grid of non-LTE model atmospheres  computed with TLUSTY and using updated model atoms of oxygen, carbon, silicon and neon. The iteration begins  with initial guesses of the stellar parameters: $T_{\rm eff}$, $\log g$,  $V \sin i$,  $\zeta$,  $\xi$,  and O and Si abundances (initially adopted as the solar values obtained by Asplund et al. 2009). The ionization balance of silicon (based on \ion{Si}{2} lines      $\lambda$  4128, 4130, 6347 and 6371 \AA\  and \ion{Si}{3})  lines $\lambda$ 4552, 4567, 4574 and 4716  \AA) is used to constrain the $T_{\rm eff}$, while the wings of the Hydrogen lines H$\gamma$ and H$\delta$ are used to define $\log g$. The selected absorption profiles are fitted independently and the best fits between synthesis and observations are obtained by $\chi^2$  minimizations. The silicon abundances are derived for a range of  microturbulent velocity values, and the selected $\xi$ is the one that  corresponds to the lowest dispersion in the abundance distribution. $V \sin i$ is allowed to vary in order to obtain the best fit for every profile and $\zeta$ is included as an additional broadening parameter needed to fit the wings of metal lines.  Once the stellar parameters are defined, one more iteration produces the final abundances of oxygen and silicon. }

{
For the rest of our sample, which present shallower and blended lines due to higher values of $V \sin i$, the iterative method described above, which was used for the sharp lined stars, could not be directly applied. We then determined the parameters of these stars based on best fits of four spectral regions: $\lambda\lambda$ 4630-4665 \AA, containing  CNO lines; $\lambda\lambda$ 4280-4410 \AA, containing the $H\gamma$ line; $\lambda\lambda$ 4450-4500 \AA, containing  \ion{He}{1} 4471 \AA\ line, and  $\lambda\lambda$ 4540-4585 \AA, containing  the \ion{Si}{3} triplet at 4552 - 4575 \AA. For obtaining the best fits between models and observations we adopted the average Si and O abundances derived for the benchmark stars.}

{ 
The procedure used was as follows: 1) We started with the initial $\log g$ and $T_{\rm eff}$ from \citet{2006Dufton} 2) We fit the CNO region near 4650 \AA\ to adjust $T_{\rm eff}$, while keeping $\log g$ fixed. 3) We adjusted $\log g$ to fit the H$\gamma$ line while keeping $T_{\rm eff}$ fixed.  4) We again adjusted $T_{\rm eff}$ to fit the 4471\,\AA\ \ion{He}{1} line while keeping $\log g$ fixed. 5) Using the $\log g$ from step 3 and the average of the $T_{\rm eff}$ values from steps 2 and 4, we fitted the \ion{Si}{3} triplet while keeping the Si abundance fixed at the mean abundance found for the four benchmark stars and varying the microturbulent velocity.  6) Finally, for sample stars with $T_{\rm eff} \gtrapprox$24\,000~K, we checked the fit to \ion{He}{2} 4686 \AA\ to verify the final parameters. The lines of the rapidly rotating stars are dominated by rotational  broadening, making it difficult to constrain the macroturbulent velocity.  We fixed $\zeta$ to 15 km/s for these stars, except for ESL 04, which required $\zeta$=30 km/s in order to obtain better fits to the profiles. In addition, $V \sin i$ values were allowed to vary in order to obtain the best fit for the \ion{Si}{3} triplet and the blended region at 4650 \AA.
}

The resultant parameters are listed in Table \ref{tab:optical} and, for ease of comparison, we also repeat the parameters from Table \ref{tab:prevwork} derived by \citet{Morel2022} where they are available, and from \cite{Hunter2009} or \citet{2006Dufton} otherwise. 
We find that the present effective temperature and surface gravity estimates are higher than in \citet{2006Dufton}, especially for the hotter stars classified as B1 or B1.5.  
This is consistent with \citet{2016ApJ...824....3P} who argued that these higher values, derived from the \ion{Si}{2}/\ion{Si}{3} ionization balance, are to be preferred to the lower values, that were derived from the \ion{Si}{3}/\ion{Si}{4} ionization balance. {\bf After excluding ESL 25, for which we could not fit the UV spectrum, for the remaining nine stars in common with \citet{Morel2022}, agreement is generally quite good, with a mean difference, (this paper minus Morel), in temperature of $+368\pm623$\ K}, and mean differences in $\log g$\ of less than 0.1 dex. Note that neither \citet{2006Dufton}, nor\ \citet{Hunter2009}, nor \citet{Morel2022}, included macroturbulent broadening in their analysis. As a result, for several of the more luminous and slower rotating stars where we have included significant macro-turbulent broadening in our fits, we find noticeably smaller projected rotational velocities than was found previously.  However, for the majority of the stars, macroturbulence does not make a  significant contribution to the line profiles.

\begin{deluxetable}{ccccccccccc}
\tablecaption{Stellar Parameters from Optical Spectra\label{tab:optical}}
\tablehead{\colhead{NGC 3293} &
\multicolumn{8}{c}{This work} & \multicolumn{2}{c}{Independent Results} \\
           \colhead{ESL No.} & \colhead{$T_{ef\!f}$} & \colhead{$\log\,g$} & \colhead{$\xi$} & \colhead{$v\sin i$} & \colhead{$\zeta$} & \colhead{RV} &\colhead{A(Si)}& \colhead{A(O)} &  \colhead{$T_{ef\!f}$} & \colhead{$\log\,g$} }
\startdata
   002\tablenotemark{a} & $24200 \pm 1000 $& $ 3.00 \pm 0.10 $ &   14   &     65  &   42 &   $-12$  &$7.54\pm0.02$ &$8.65\pm0.07$ & 23200\tablenotemark{b}  &2.80   \\
   003\tablenotemark{a} & $22500 \pm 1000 $& $ 3.00 \pm 0.10 $ &   12   &     54  &   27 &   $-14$  &$7.41\pm0.05$ &$8.68\pm0.08$ & 20500\tablenotemark{b}  &2.75   \\
   004 & $24000 \pm 1000 $& $ 3.04 \pm 0.10 $ &   15   &     95  &   30 &   $-20$  & mean & mean &                 22700\tablenotemark{b}  &3.13   \\
   005 & $22700 \pm 1500 $& $ 3.00 \pm 0.15 $ &   12   &     180 &   15 &   $-18$  & mean & mean &                 21500\tablenotemark{c}  & 3.05 \\
   006 & $24000 \pm 1500 $& $ 3.15 \pm 0.15 $ &   15   &     200 &   15 &   $-18$  & mean & mean &                 21500\tablenotemark{c}  & 3.15  \\
   007\tablenotemark{a} & $23150 \pm 1000 $& $ 3.08 \pm 0.10 $ &   12   &     28  &   26 &   $-17$  &$7.52\pm0.08$ &$8.75\pm0.10$ &22600\tablenotemark{b}  & 3.25   \\
   008 & $23600 \pm 1000 $& $ 3.40 \pm 0.10 $ &   12   &    130  &   15 &   $+55$  & mean & mean &                 21500\tablenotemark{c}  &  3.25  \\
   010\tablenotemark{a} & $23600 \pm 1000 $& $ 3.55 \pm 0.10 $ &   11   &     35  &   30 &   $-18$  &$7.44\pm0.05$ &$8.60\pm0.05$ &24493\tablenotemark{d}   & 3.61     \\    
   012 & $24500 \pm 1000 $& $ 3.70 \pm 0.10 $ &   12   &    90   &   15 &   $-25$  & mean & mean &                24444\tablenotemark{d}   & 3.70   \\   
   015 & $26000 \pm 1500 $& $ 3.80 \pm 0.15 $ &    6   &    260  &   15 &   $-10$  & mean & mean &                25000\tablenotemark{c}   &3.80       \\   
   016 & $25300 \pm 1000 $& $ 3.80 \pm 0.10 $ &   10   &     55  &   15 &   $-32$   & mean & mean &     \nodata     &  \nodata\\
   019 & $25500 \pm 1000 $& $ 3.75 \pm 0.10 $ &    8   &    120  &   15 &   $-20$   & mean & mean &               24380\tablenotemark{d}   & 3.63   \\ 
   023 & $22700 \pm 1000 $& $ 3.90 \pm 0.10 $ &    5   &    170  &   15 &   $-18$   & mean & mean &               21730\tablenotemark{d}   & 3.71   \\  
   024 & $22200 \pm 1000 $& $ 3.85 \pm 0.10 $ &    5   &    160  &   15 &   $-18$   & mean & mean &               21700\tablenotemark{d}   & 3.69   \\  
   025 & $21500 \pm 1500 $& $ 3.70 \pm 0.15 $ &    5   &    215  &   15 &   $-15$   & mean & mean &               20460\tablenotemark{d}   & 3.69   \\  
   028 & $21200 \pm 1500 $& $ 3.90 \pm 0.15 $ &    4   &    200  &   15 &   $-10$   & mean & mean &               20429\tablenotemark{d}   & 3.86   \\  
   030 & $19800 \pm 1500 $& $ 3.80 \pm 0.15 $ &    6   &    220  &   15 &   $-8$    & mean & mean &               19675\tablenotemark{d}   & 3.80   \\   
   031 & $21000 \pm 1500 $& $ 4.00 \pm 0.15 $ &    6   &    230  &   15 &   $-8$    & mean & mean &               20322\tablenotemark{d}   & 3.86   \\   
   038 & $20200 \pm 1500 $& $ 3.95 \pm 0.15 $ &    6   &    235  &   15 &   $-15$   & mean & mean &               20216\tablenotemark{d}   & 3.94   \\  
\enddata
\tablenotetext{a}{Stars fit using the full procedure of \citet{Bragancga2019}. }
\tablenotetext{b}{\citet{Hunter2009}}
\tablenotetext{c}{\citet{2006Dufton}}
\tablenotetext{d}{\citet{Morel2022}}
\end{deluxetable}

\begin{figure}
    \centering
    \includegraphics[width=\linewidth]{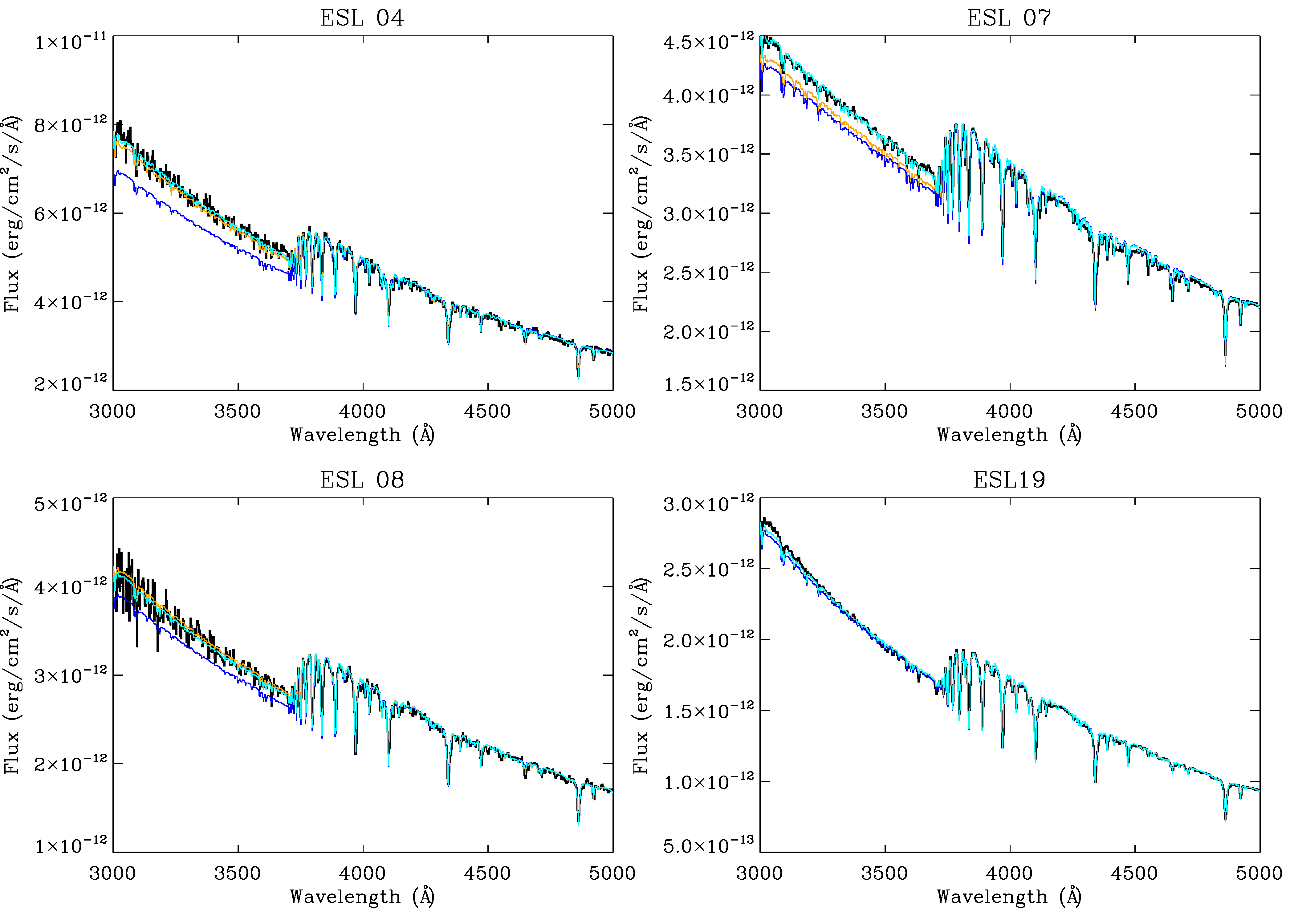}
    \caption{SEDs from G430L data (black) and comparison with the best fitting model (cyan). Also shown are the model SEDs for the parameters derived from the high resolution optical data in the present work (orange) and from \citet{2006Dufton} (blue). }
    \label{fig:seds}
\end{figure}

Four of our sources, ESL 04, 07, 08, and 19, have HST G430L spectrophotometry \citep{2019ApJ...886..108F} covering the important Balmer jump region, that is sensitive to both effective temperature and surface gravity.  These data were used to check the present solutions for these stars by finding the model that best reproduces the G430L spectral energy distribution (SED). 
We used the Levenberg-Marquardt algorithm, \citep{2009ASPC..411..251M}, and a grid of TLUSTY models for each star, normalized to the observed spectrum in the $V$-band, with extinction applied according to the observed and model $B-V$ measurements and appropriate rotational and radial velocity corrections, and then re-binned to the G430L resolution and pixel size.  The results are illustrated in Figure \ref{fig:seds} and summarized in Table \ref{tab:seds}.

Inspection of Figure \ref{fig:seds} confirms that the parameters derived from our spectroscopic analysis of the high resolution optical data also provide an excellent match to observed SEDs, except for star ESL\,007. 
However close inspection of the Balmer jump of that star reveals the presence of a weak secondary emission jump, suggesting this is a Be star \citep{2020A&A...634A..18C}, an hypothesis supported by the presence of previously overlooked weak emission in the core of the H$\alpha$ profile. 
While weak narrow nebular H$\alpha$ emission is present for most of the sources discussed here, that of ESL\,007 is clearly broader than nebular emission. 

The spectroscopic parameters are therefore to be preferred over the SED results for ESL\,007. 
Comparing results for the other three stars it is clear that the higher effective temperatures derived for the B1/B1.5 stars using the present spectroscopic and SED approaches are indeed appropriate for this spectral type.

\subsection{Luminosities}
We adopt a distance to NGC\,3293 of 2335$^{+58}_{-56}$\,pc that is based on an analysis of $Gaia$ DR3 data for high probability cluster members (Molina Lera et al. in preparation), in good agreement with the $Gaia$ DR2 distance of 2334$\pm$52\,pc \citep{2021MNRAS.504..356D}.
We note the distance to NGC\,3293 is similar to other major stellar OB groups in the Carina nebula, such as Tr\,14 and 16 \citep{2022A&A...657A.131M}, and all are consistent with being at the geometric distance of 2350$\pm$50 pc derived from an analysis of the 3D structure of the homunculus around $\eta$ Car \citep{2006ApJ...644.1151S}.
Using the above distance and effective temperatures (from Table \ref{tab:optical}), we calculate stellar radii and luminosities, taking into account bolometric corrections (BC) for the TLUSTY models \citep{2007ApJS..169...83L} and extinction.
We adopt the average Galactic extinction law of \citet{2019ApJ...886..108F} (that assumes $R_{V}$=3.1 following \citet{2003A&A...402..549B}) to redden the model SEDs and estimate $B-V$, $E(B-V)$, $A_{V}$ and $M_{V}$ via comparison with the observed $BV$ magnitudes (Table \ref{tab:LandM}).

\begin{deluxetable}{ccccccccccc}
\tablecaption{Comparison of stellar parameters for the subsample with HST G430L spectrophotmetry (SED) with spectroscopic results from the present work and from the literature, as indicated. SED errors are formal 1$\sigma$ errors in the least squares fit to the data.\label{tab:seds}}
\tablehead{\colhead{NGC 3293} &
\multicolumn{2}{c}{This work} & \multicolumn{2}{c}{\citet{2006Dufton}} & \multicolumn{2}{c}{\citet{2016ApJ...824....3P}}  & \multicolumn{2}{c}{SED} & \multicolumn{2}{c}{\citet{2019ApJ...886..108F}} \\
           \colhead{ESL No.} & \colhead{$T_{ef\!f}$} & \colhead{$\log\,g$} & \colhead{$T_{ef\!f}$} & \colhead{$\log\,g$} & \colhead{$T_{ef\!f}$} & \colhead{$\log\,g$} & \colhead{$T_{ef\!f}$} & \colhead{$\log\,g$} & \colhead{$T_{ef\!f}$} & \colhead{$\log\,g$} }
\startdata
   004 & 24000 &  3.04 & 21500$\pm$1000 & 3.00$\pm$0.10 & 22630$\pm$1500 &  3.04$\pm$0.25 & 25944$\pm$290 & 3.33$\pm$0.06 & 25037 & 3.10 \\
   007 & 23150 &  3.08 & 22800$\pm$1000 & 3.10$\pm$0.10 & 22474$\pm$1500 &  3.07$\pm$0.25 & 24947$\pm$921 & 3.17$\pm$0.15 & 24529 & 3.25 \\
   008 & 23600 &  3.40 & 21500$\pm$1000 & 3.25$\pm$0.10 & 22604$\pm$1500 &  3.50$\pm$0.25 & 23319$\pm$224 & 3.43$\pm$0.07 & 24490 & 3.36 \\
   019 & 25500 &  3.75 & 25000$\pm$1000 & 3.85$\pm$0.10 & 24689$\pm$1500 &  3.85$\pm$0.25 & 25512$\pm$266 & 3.92$\pm$0.09 & 25921 & 3.71 \\
\enddata
\end{deluxetable}

Given the brightness of these stars, it is to be expected that uncertainties in the apparent magnitudes are dominated by systematics and possible variability.
While Johnson photometry with notional 1$\sigma$ errors of 0.03$^m$ has been published by \cite{2005A&A...437..467E} and \citet{2003A&A...402..549B}, there are some clear outliers among the sources in common, with some differences ranging up to approximately 3$\sigma$ in both $V$ and $B-V$. 
We therefore compared these datasets with the photometry derived from $Gaia$ XP spectra, that exists for 17 of our sources, and with the photometry derived from the 4 sources with HST spectrophotometry discussed above.
We followed the procedure outlined in \citet{2007ASPC..364..227M}, and used the zero-points noted therein. Excellent agreement was found between the $Gaia$ and HST results, with $V_{gaia}-V_{hst}\sim -0.02$ and $(B-V)_{gaia}-(B-V)_{hst}\sim-0.0195$.  
However the $Gaia$ magnitudes are consistently brighter and bluer than published data by up to 0.1$^m$ in $V$ and $B-V$. 
We suspect that some of these differences may be attributed to variability (many sources are $\beta$ Cepheid variables), and the published data are single epoch measurements, while the $Gaia$ spectra were taken over multiple epochs. 
The photometry derived from the $Gaia$ XP spectra is therefore adopted in the present work, with uncertainties of 0.02$^m$ being used for both $V$ and $B-V$. 
Stars ESL\,028 and 030 do not have XP spectra, for these we used transformation formulae \citep{2021A&A...649A...3R, 2000PASP..112..961B} to convert $Gaia$ native photometry to the $V$ and $B-V$, adopting slightly larger uncertainties of 0.03$^m$ for these values.

\begin{deluxetable}{cccccccccc}
\tablecaption{Observed and derived stellar parameters.  \label{tab:LandM} }
\tablehead{\colhead{ESL} &
           \colhead{$V$} &
           \colhead{$B-V$}&
           \colhead{$E(B-V)$}&
           \colhead{$T_{\rm eff}$}&
           \colhead{$\log g$}&
           \colhead{$M_V$}&
           \colhead{BC}&
           \colhead{$\log L/L_\odot$}&
           \colhead{$R / R_\odot$}}
\startdata
  2 &  6.67 &  0.03 & 0.259 & 24200$\pm$1000 & 3.00$\pm$0.10 & $-5.97$ & $-2.346$ & 5.224$\pm$0.054 & 23.39$\pm$1.20     \\
  3 &  7.58 &  0.05 & 0.271 & 22500$\pm$1000 & 3.00$\pm$0.10 & $-5.10$ & $-2.180$ & 4.809$\pm$0.057 & 16.79$\pm$0.88     \\
  4 &  7.99 & $-0.01$ & 0.220 & 24000$\pm$1000 & 3.04$\pm$0.10 & $-4.53$ & $-2.329$ & 4.640$\pm$0.055 & 12.14$\pm$0.62   \\
  5 &  8.08 &  0.04 & 0.261 & 22600$\pm$1500 & 3.00$\pm$0.15 & $-4.57$ & $-2.186$ & 4.600$\pm$0.077 & 13.12$\pm$0.89     \\
  6 &  8.18 &  0.02 & 0.252 & 24000$\pm$1500 & 3.15$\pm$0.15 & $-4.44$ & $-2.334$ & 4.604$\pm$0.074 & 11.70$\pm$0.76     \\
  7 &  8.21 &  0.13 & 0.356 & 23150$\pm$1000 & 3.08$\pm$0.10 & $-4.74$ & $-2.250$ & 4.690$\pm$0.056 & 13.83$\pm$0.72     \\
  8 &  8.54 &  0.00 & 0.232 & 23600$\pm$1000 & 3.40$\pm$0.10 & $-4.02$ & $-2.312$ & 4.428$\pm$0.055 &  9.84$\pm$0.49     \\
 10 &  8.70 & $-0.05$ & 0.182 & 23600$\pm$1000 & 3.55$\pm$0.10 & $-3.71$ & $-2.321$ & 4.306$\pm$0.057 &  8.56$\pm$0.43   \\
 12 &  8.91 &  0.02 & 0.258 & 24500$\pm$1000 & 3.70$\pm$0.10 & $-3.73$ & $-2.417$ & 4.354$\pm$0.056 &  8.38$\pm$0.41     \\
 15 &  9.09 & $-0.06$ & 0.186 & 26000$\pm$1500 & 3.80$\pm$0.15 & $-3.33$ & $-2.556$ & 4.250$\pm$0.072 &  6.61$\pm$0.38   \\
 16 &  9.31 & $-0.02$ & 0.222 & 25300$\pm$1000 & 3.80$\pm$0.10 & $-3.22$ & $-2.496$ & 4.182$\pm$0.055 &  6.45$\pm$0.31   \\
 19 &  9.22 & $-0.08$ & 0.163 & 25500$\pm$1000 & 3.75$\pm$0.10 & $-3.13$ & $-2.510$ & 4.151$\pm$0.054 &  6.13$\pm$0.29   \\
 23 &  9.95 & $-0.07$ & 0.155 & 22700$\pm$1000 & 3.90$\pm$0.10 & $-2.37$ & $-2.256$ & 3.747$\pm$0.058 &  4.85$\pm$0.24   \\
 24 &  9.95 & $-0.05$ & 0.172 & 22200$\pm$1000 & 3.85$\pm$0.10 & $-2.42$ & $-2.200$ & 3.746$\pm$0.060 &  5.07$\pm$0.25   \\
 25 & 10.14 & $-0.04$ & 0.179 & 21500$\pm$1500 & 3.70$\pm$0.15 & $-2.25$ & $-2.118$ & 3.645$\pm$0.087 &  4.81$\pm$0.30   \\
 28 & 10.24 &  0.01 & 0.223 & 21200$\pm$1500 & 3.90$\pm$0.15 & $-2.29$ & $-2.093$ & 3.650$\pm$0.096 &  4.99$\pm$0.35     \\
 30 & 10.51 & $-0.03$ & 0.174 & 19800$\pm$1500 & 3.80$\pm$0.15 & $-1.87$ & $-1.924$ & 3.414$\pm$0.101 &  4.35$\pm$0.30   \\
 31 & 10.59 &  0.02 & 0.231 & 21000$\pm$1500 & 4.00$\pm$0.15 & $-1.97$ & $-2.075$ & 3.513$\pm$0.091 &  4.33$\pm$0.26     \\
 38 & 10.95 & $-0.01$ & 0.195 & 20200$\pm$1500 & 3.95$\pm$0.15 & $-1.50$ & $-1.980$ & 3.286$\pm$0.097 &  3.60$\pm$0.21   \\
\enddata
\end{deluxetable}

\begin{figure}
    \centering
    \includegraphics[width=\linewidth]{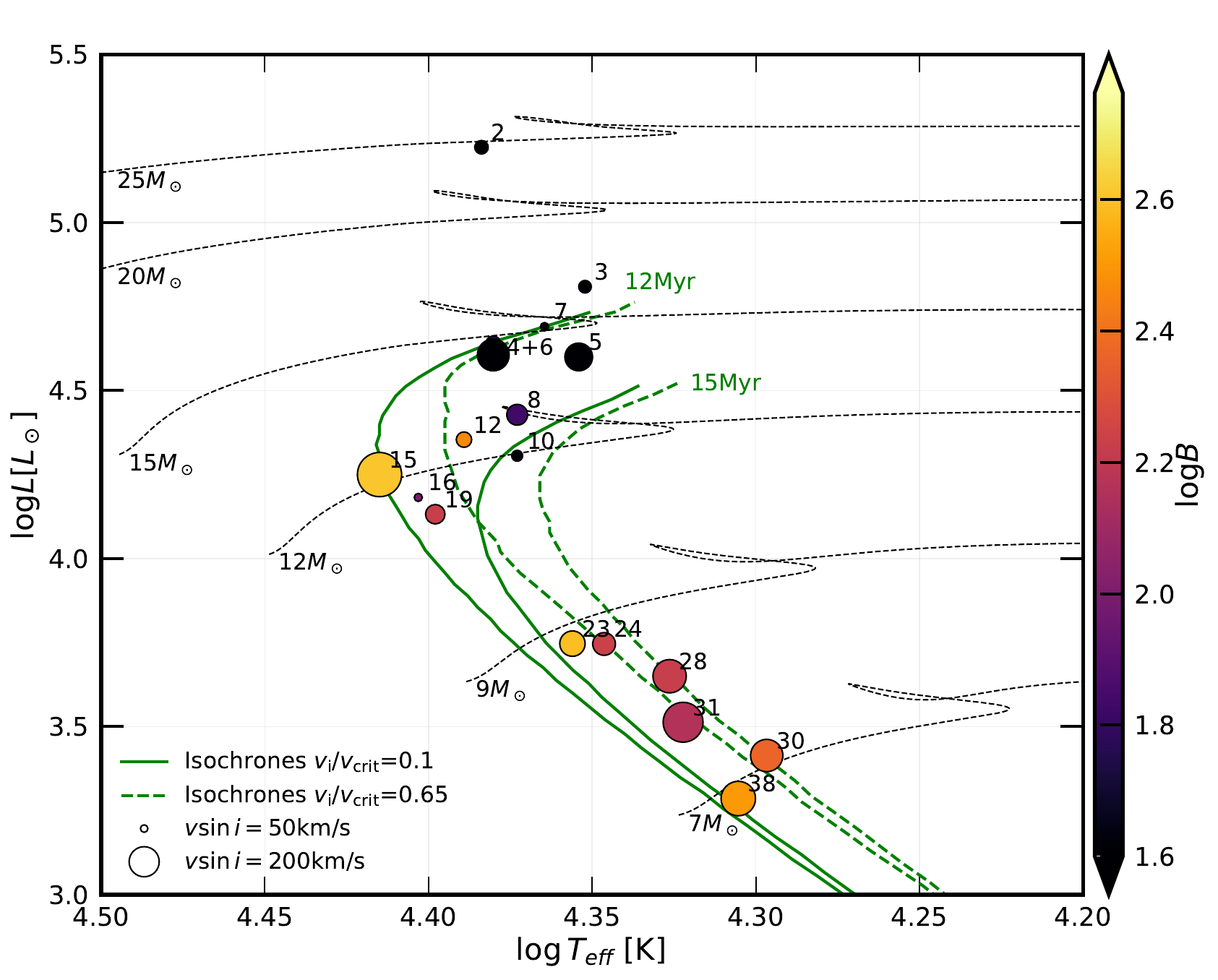}
    \caption{Location of stars of NGC\,3293 with boron measurements in the HR-diagram. Stars are shown as circles, which increase in size proportional to $v \sin i$, and are filled with colors that correspond to their surface boron abundance. Stars with upper limits on their surface boron abundance are colored in black.
    Evolutionary tracks (black dashed lines) are from non-rotating stellar models, and the isochrones (green lines) from rotating models of \citep{Jin2023a},  with $v_\mathrm{i}/v_\mathrm{crit}=0.1$ and $0.65$, and for ages of 12\,Myr and 15\,Myr.}
    \label{fig:hrd}
\end{figure}

In Figure \ref{fig:hrd}, we show the resulting locations of these stars in the HR diagram together with selected evolutionary tracks from \citet{Jin2023a}. The physics they adopted, including the assumed efficiencies for rotationally driven mixing are similar to those of the model \citet{2011A&A...530A.115B, 2011A&A...530A.116B}, but the new \citet{Jin2023a} calculations were made using the MESA code \citep{2011ApJS..192....3P, 2013ApJS..208....4P, 2015ApJS..220...15P, 2018ApJS..234...34P, 2019ApJS..243...10P} version 10398, and also included some updates to the adopted input abundances, opacities, and mass-dependent overshooting.  

Traditional isochrone fitting done using non-rotating models has yielded a range of ages for this cluster. E.g., \citet{2003A&A...402..549B} estimated $8.1 \pm 1\,$Myr from the apparent nuclear age of the turnoff and $\approx 10$ Myr from the luminosity of the pre-main sequence stars relative to the zero age main-sequence based on a significantly larger distance estimate of 2750\,pc, while more recently \citet{2021MNRAS.503.5929B} found an age of $12 \pm 3\,$Myr.  { \citet{Morel2022}} included estimates of the effects of rotational distortion and mixing and suggested a significantly older age near 20 Myr.  However, this result will be sensitive to how the adopted details of the interior mixing affect stellar lifetime, and still assumed that most observed stars in this cluster could be explained as products of single star evolution. 

In recent years it has become apparent that young star clusters cannot be treated as a single population that can be modeled by one isochrone, or perhaps not even by a mix of different ages and rotation rates, (e.g., \citet{2007MNRAS.379..151M}). Instead, most young clusters in the Magellanic clouds show an extended main-sequence (eMSTO) and a split of the main-sequence into a blue and a red sequence \citep{2023A&A...672A.161M}, with the blue sequence predominantly consisting of relatively narrow-lined stars and the red one of more rapid rotators \citep{2017ApJ...846L...1D, 2018AJ....156..116M}. Similar features are seen in Galactic clusters, \citep{2018MNRAS.480.3739B, 2018ApJ...869..139C}.  Suggestions for the origin of the two rotational distributions include stellar mergers \citep{2022NatAs...6..480W}, which are expected to lead to slowly rotating merged products \citep{2019Natur.574..211S}, or differences in the lifetime of the pre-main sequence accretion disk \citep{2020MNRAS.495.1978B}, with stars that retain their disks longer having smaller rotational velocities.

In NGC 3293, most of the stars considered by \citet{2016ApJ...824....3P}, as well as the brighter ones in this paper may well correspond to eMSTO stars, and as such may well be the products of binary interaction. The fainter stars, ESL 23, 24, 28, 30, 31, and 38, would in this interpretation be part of the red main-sequence due to their fast rotation and can be treated as stars that have evolved as single rotating stars without effects from mergers or binary interactions. The eMSTO stars might be expected to be depleted in boron, especially if these are mass transfer remnants { \citep[see fig.\,9 in][]{2012ARA&A..50..107L, 2001A&A...369..939W}}, while the red sequence stars should show more modest boron depletion due to the effects of rotationally driven mixing.

\section{Fitting of UV Spectra}

Our procedures for fitting the UV spectra are very similar to those used in our previous work \citep{2016ApJ...824....3P}.  For the adopted $T_{\rm eff}$ and $\log g$ values, we refit the spectrum allowing the boron abundance and radial velocity to vary, and to account for imperfect modeling of the spectral resolution and line formation, we also allow the micro-turbulence and rotational velocity to vary from the values determined by the best fit to the optical data. At each step of the iterative solution, we interpolate in the grid of TLUSTY models to the desired parameters, and recalculate the synthetic spectrum. The Levenberg-Marquardt algorithm \citep{2009ASPC..411..251M}, was then used to find the best fit. 

{ As the \ion{B}{3} feature is heavily blended with lines of iron group elements, before fitting this line, it is essential to obtain the best possible fit to the line profiles of the iron group elements. In practice, this requires fine tuning the micro-turbulence, and, especially for the lower mass stars in our sample, the value of $\xi$\ implied by the UV spectra is often substantially lower than the value determined from the optical data.  However, if this lower $\xi$\  is actually compensating for uncertainties in the non-LTE model of the iron ionization equilibrium, then one might argue that the higher, optically determined values are more appropriate for the synthesis of the boron feature itself.  While the need to fit the overall UV spectrum precludes us from getting a useful overall fit with the higher  $\xi$ value, we can easily determine the change in abundance needed to yield the same equivalent width for the boron feature with the alternate micro-turbulence.  Even for ESL 38, where the optically determined value of $\xi = 6\,$km/s is much larger than the $\approx 1\,$km/s needed for the UV, the implied change in the boron abundances is only $-0.19\,$dex. For ESL 19, 23, 24, 30, and 31 the differences range from $-0.06$ to $-0.15\,$dex with a mean difference of about $-0.1\,$ dex.  For the boron feature, the effect of vmicro uncertainties is somewhat muted not only because these lines are relatively weak, but also because the thermal broadening of the boron line is relatively large (at 20000K, $(V_{therm}\approx 5.5\,$km/s for boron vs $2.4\,$km/s for iron), and the boron line is additionally split by the isotopic and hyperfine structure, which further reduces the saturation at small values of $\xi$.
}

To estimate other contributions to the final errors, we perturb a number of our assumptions and then redetermine the best fit boron abundance for each case.  Our detailed procedure is as follows:

\begin{enumerate}
\item Correlated error in $T_{\rm eff}$ and $\log g$:  These parameters are tightly correlated, as adopting a higher $T_{\rm eff}$ requires a higher $\log g$ to fit the
Balmer lines. So we will perturb these together using the best fit correlation between Teff and logg as determined from fitting the H-$\gamma$\ 
line.  The adopted temperature error is either 1000 or 1500K (see table /ref{tab:optical}), while the corresponding gravity variation varies from 0.1 to 0.15 dex per 1000K, similar to the quoted errors in $\log g$.

\item Error in $\log g$ at fixed $T_{\rm eff}$: We also calculate the change in boron abundance for a 0.05 dex change in $\log g$ at fixed $T_{\rm eff}$.

\item  Formal fitting error:  The Marquardt-Levenburg algorithm returns an error estimate for each variable.  By refitting just the boron
line with all other parameters frozen, we can get an estimate of the expected errors due to the S/N of the data. For most of our stars, these errors are in the range of 2 to 5\%.

\item Uncertainties in line synthesis:  We do an alternate set of calculations with the modified line parameters for blending lines discussed below, and take
the difference between the boron abundance derived with the two line sets as an additional error term.  For the more rapidly rotating stars in this sample, the impact of blending lines can be larger than was the case for the stars analyzed by \cite{2016ApJ...824....3P}, and so we reconsidered the lines close to the boron feature (Table \ref{tab:linedata}) to ensure that we were not underestimating the opacity in these blending features. Most of the opacity in the blending lines just shortward of the boron feature is from  \ion{Fe}{3} lines which are treated in NLTE in the TLUSTY models. However, the opacity immediately longward of the boron line is mostly due to \ion{Mn}{3} with a \ion{Ni}{3} line also contributing.  The Mn and Ni lines are treated only in LTE in the TLUSTY models and this may introduce errors that vary as a function of $T_{\rm eff}$ and $\log g$ due to differential NLTE effects that we are not accounting for.
For the \ion{Mn}{3} line at 2066.546\,\AA, we adopted the nominal transition probability \citep{K10}, although previous work, \citep{1999ApJ...516..342P}, had suggested that a value lower by 0.25 to 0.28 dex might be more appropriate.  
We have also increased the opacity in the blend of \ion{Fe}{3}\ $\lambda2066.144$\,\AA\  (vacuum wavelength) and \ion{Mn}{3}\ $\lambda2066.160$\AA\ lines just shortward of the \ion{B}{3}\ 2066.4\,\AA\ line. 
In the narrow lined template stars that \citet{2016ApJ...824....3P} used to fine-tune the line list, these lines are normally blended with the ISM \ion{Cr}{2} line at 2066.164\,\AA, and so are difficult to evaluate empirically, but a reconsideration of the high S/N spectrum of HD 36285 led us to also increase the opacity in this blend.

\item Errors in continuum normalization:  To estimate the effects of errors in the flat fielding of the spectra, we perturb the overall continuum normalization by 0.5\% while leaving all other parameters fixed. For the broad-lined stars, especially at the lower masses, this term is often the dominant one in the error analysis.  

\end{enumerate}

All of the above errors are then added in quadrature to derive the final error estimate. Upper limits are a bit more subjective, but
include consideration of the perturbed model that found the highest boron abundance and the estimated errors for that model. Note that
for ESL006, all cases yield valid fits with non-zero abundances, but many of these aren't significantly different from the
no-boron case and so we've assigned an upper limit for this one as well, even though our procedure formally finds a NLTE corrected boron abundance of $1.17 \pm 0.44$ for this star.

We note that the overall rms variation in our measured abundances for ESL 23, 24, 28, 30, 31, and 38 is only about 0.16 dex, while our mean estimated "error" for these stars is 0.23 dex, This suggests that our normalization and error estimation procedures are at least consistent from star-to-star, although this leaves open the possibility of a larger systematic error that affects all of these stars uniformly.

\begin{deluxetable}{cclrrl}
\tablecaption{Strong lines near 2066 \AA.\label{tab:linedata}}
\tablehead{\colhead{$\lambda$ (vac)} & \colhead{$\lambda$ (air)} & \colhead{Ion} & \colhead{$\log gf$} & E (lower) & \colhead{comments} \\
           \colhead{\AA}              & \colhead{\AA}             &               &                     & cm$^{-1}$    &}
\startdata
2065.880 & 2065.220 &  Fe  III &  -1.55  &  89697.471   &        \\                                      
2065.898 & 2065.238 &  Fe  III &  -2.47  &  84159.501   &         \\                                     
2065.935 & 2065.275 &  Fe  III &   0.00  & 130856.736   &               \\                               
2065.936 & 2065.276 &  Fe  III &  -0.77  &  89783.551   &                  \\                            
2066.144 & 2065.484 &  Fe  III &  -0.17  & 130861.645   &  Scaled to match HD36285        \\             
2066.160 & 2065.500 &  Mn  III &  -0.29  &  94697.847   &  Scaled to match HD36285               \\      
2066.430 & 2065.771 & $^{11}$B III  & -0.413  & 0.000     & Mean of hyperfine, $gf$ scaled by 0.8\\
2066.115 & 2065.455 & $^{10}$B III  & -1.014  & 0.000     & Mean of hyperfine, $gf$ scaled by 0.2\\
2066.546 & 2065.886 &  Mn  III &   0.01  &  85346.719   &                                  \\        
2066.824 & 2066.164 &  Ni  III &  -0.95  &  85834.200   &  Scaled to match HD36285           \\          
2067.032 & 2066.372 &  Mn  III &   0.57  &  85077.091   &                                \\              
2067.084 & 2066.424 &  Ni  III &  -0.59  &  86645.880   &                               \\               
2067.101 & 2066.441 &  Mn  III &  -0.16  &  94771.466   &                                              
\enddata
\end{deluxetable}

The quality of the fits we obtain are very similar to those found in \citet{2016ApJ...824....3P}. However, the abundances are generally somewhat smaller due to the increases in line opacity discussed above.  For the cooler, higher gravity stars where the boron abundance is relatively large, these changes are largely offset by the somewhat lower effective temperatures we have adopted here relative to \citet{2016ApJ...824....3P}.  For the partially depleted stars the differences are somewhat larger.

Our final adopted boron abundances and uncertainties are shown in Table \ref{tab:uvresults}, and for the stars in common  these are compared to the values of \citet{2016ApJ...824....3P}.  
While the spectral synthesis of boron is also done in LTE as the TLUSTY models do not include this model ion, we estimate the NLTE correction using the results of \citet{2002ApJ...565..571V}.
In interpreting these results, it is important to consider other issues not considered in the formal fits that can potentially effect the derived abundances.  

In particular, throughout this analysis we neglect the impact of the rotational distortion of the star on the emitted spectrum, which will in fact depend on rotational velocity and aspect angle, $\sin i$. For example, our treatment of rotational broadening uses a simple broadening function that conserves equivalent width, hence it's natural to wonder to what extent this compromises our results. However, \citet{collins1991} has performed detailed LTE calculations of the continua, H, He and \ion{Mg}{2} line profiles for a grid of B-type stars (B1 to B9 dwarfs and giants) with various rotational velocities and $\sin i$ values. These results demonstrate that there is significant impact on the line profiles only for cases where the rotational velocities greater than $\sim 80$\% critical, and are strongest at high inclination. While we cannot as yet rule out such a configuration for an individual star, we infer that our sample is likely not significantly affected by the simplified treatment of rotation.  At higher rotational rates, more detailed calculations of the effect on the relative line strengths would be needed, \citep{2005A&A...440..305F}.

\begin{deluxetable}{ccccccccccc}
\tablecaption{Results of UV Spectral fitting.}
\label{tab:uvresults}
\tablehead{
\multicolumn{8}{c}{This work} & \multicolumn{3}{c}{\citet{2016ApJ...824....3P}} \\
    \colhead{ESL \#} & \colhead{$T_{ef\!f}$} & \colhead{$\log\,g$} & \colhead{$\xi$} & \colhead{$v\sin i$} & \colhead{RV} & 
    \colhead{B (LTE)} & {B (NLTE)} & 
    \colhead{$T_{ef\!f}$} & \colhead{$\log\,g$} & {B (NLTE)}
}
    \startdata
   02 &  24200  & 3.00   &  18.49   &   89.9  & $  -15.531  $& $ <1.42  $ & $ <1.2            $  &    23200 &  2.80  & $  < 1.24        $  \\
   03 &  22500  &  3.00  &  17.20   &   83.6  & $  -14.570  $& $ <1.52  $ & $ <1.3            $  &    22100 &  2.95  & $  < 1.22        $  \\       
   04 &  24000  &  3.04  &  15.11   &   99.8  & $  -19.208  $& $  1.958 $ & $  1.743\pm 0.247 $  &    24200 &  3.30  & $  1.93\pm 0.30  $  \\
   05 &  22700  &  3.00  &  12.59   &  184.2  & $  -13.578  $& $  1.720 $ & $  1.499\pm 0.284 $  &          &        & $                $  \\      
   06 &  24000  &  3.15  &  15.96   &  211.4  & $  -11.925  $& $ <1.80  $ & $  <1.6           $  &          &        & $                $  \\   
   07 &  23150  &  3.08  &   9.99   &   56.0  & $  -20.575  $& $ <2.01  $ & $  <1.8           $  &    24100 &  3.20  & $   1.85\pm 0.74 $  \\  
   08 &  23600  &  3.40  &  10.23   &  138.2  & $   +22.974  $& $  2.015 $ & $  1.835\pm 0.263 $  &    25200 &  3.60  & $   2.41\pm 0.25 $  \\  
   10 &  23600  &  3.55  &   8.02   &   72.5  & $   -6.887  $& $  1.698 $ & $  1.533\pm 0.327 $  &    24500 &  3.50  & $   1.87\pm 0.37 $  \\  
   12 &  24500  &  3.70  &   7.29   &  102.3  & $   -5.912  $& $  2.613 $ & $  2.463\pm 0.261 $  &    23800 &  3.55  & $   2.69\pm 0.27 $  \\  
   15 &  26000  &  3.80  &   8.72   &  293.6  & $   -0.856  $& $  2.706 $ & $  2.608\pm 0.207 $  &          &        & $                $  \\    
   16 &  25300  &  3.80  &   5.64   &   54.0  & $  -42.567  $& $ <2.13  $ & $        < 2.0    $  &          &        & $                $  \\  
   19 &  25500  &  3.75  &   3.90   &  129.1  & $  -16.719  $& $  2.358 $ & $  2.223\pm 0.296 $  &    25000 &  3.85  & $   2.35\pm 0.32 $  \\  
   23 &  22700  &  3.90  &   1.48   &  168.6  & $   -9.708  $& $  2.720 $ & $  2.589\pm 0.255 $  &    24300 &  3.80  & $   2.55\pm 0.33 $  \\  
   24 &  22200  &  3.85  &   1.71   &  150.9  & $   -6.987  $& $  2.382 $ & $  2.235\pm 0.246 $  &    24300 &  3.90  & $   2.26\pm 0.34 $  \\  
   28 &  21200  &  3.90  &   1.35   &  221.5  & $   -3.610  $& $  2.402 $ & $  2.221\pm 0.215 $  &          &        & $                $  \\  
   30 &  19800  &  3.80  &   1.32   &  214.7  & $   -8.872  $& $  2.606 $ & $  2.363\pm 0.211 $  &          &        & $                $  \\  
   31 &  21000  &  4.00  &   1.38   &  265.3  & $    +5.478  $& $  2.344 $ & $  2.156\pm 0.220 $  &          &        & $                $  \\  
   38 &  20200  &  3.95  &   0.92   &  229.0  & $   -4.005  $& $  2.723 $ & $  2.502\pm 0.215 $  &          &        & $                $  \\  
\enddata
\end{deluxetable}

\begin{figure}
    \gridline{
        \fig{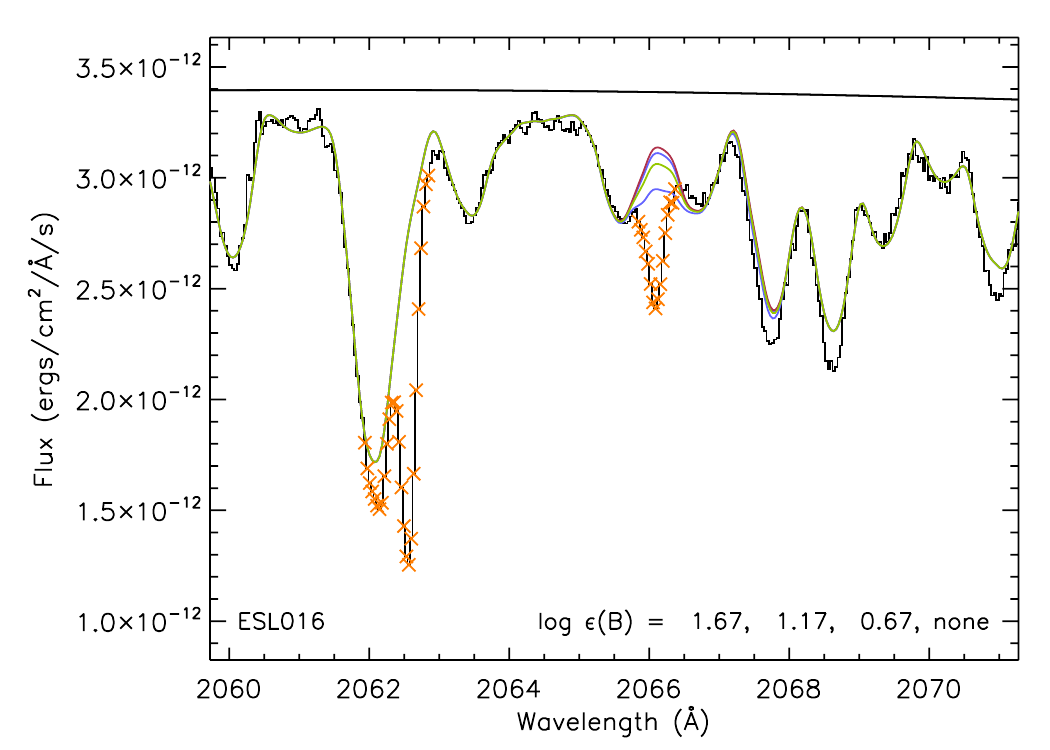}{0.49\textwidth}{(a) Using default line list}
        \fig{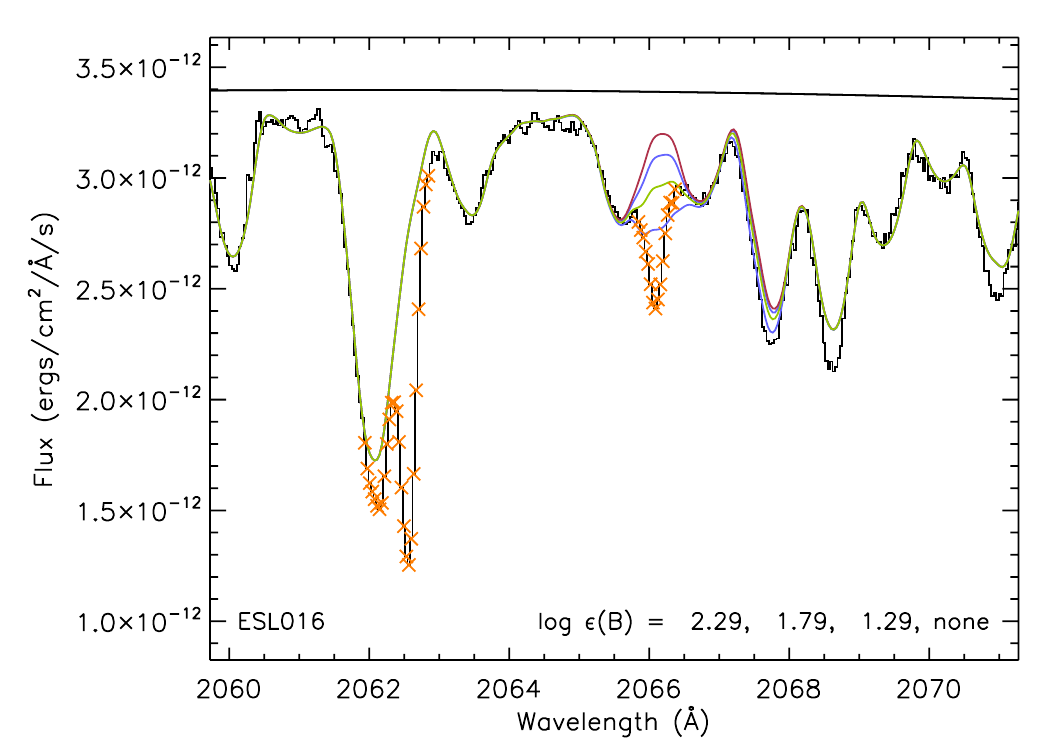}{0.49\textwidth}{(b) Using modified line list}
    }
    \caption{
    In Figs.\ \ref{fig:first-uv} through \ref{fig:last-uv}, we compare the observed spectrum for each target observed in HST program 14673 (black lines) to  synthetic spectrum constructed with a variety of assumed boron abundances (colored lines). These include the case of no boron (upper red curve), our best fit abundance or adopted upper limit (green curve), and that value $\pm0.5$\ dex (blue curves). Regions of the observed spectrum masked out of the fit due to contamination from the interstellar Zn or Cr lines are marked with orange crosses. The normalized model continuum flux is shown by the upper black line.
    For the narrow-lined stars ESL007 and ESL016, most of the equivalent width of the boron feature is obscured by the ISM \ion{Cr}{2} line. This makes the resulting fit for these stars very sensitive to minor changes in the spectral synthesis. In this figure we show the effect on the fitting of ESL 16 of the reduction in the opacity of the \ion{Mn}{3} and \ion{Ni}{3} lines discussed in the text.}
\label{fig:first-uv}
\end{figure}

Even in the absence of significant gravity darkening, there are other uncertainties to consider. For the relatively narrow-lined star ESL 07 and ESL 16, the \ion{Cr}{2} ISM feature obscures most of the equivalent width of the stellar boron feature. While in principle one could attempt to fit a model of the ISM to the three members of the \ion{Cr}{2} triplet that are covered by our data, in practice the COS G185M does not quite resolve the individual components, the strength of which appear to vary significantly and independently between site-lines in this cluster. In addition, the other two components of the triplet fall in regions where the stellar opacity is changing rapidly introducing additional uncertainties. Given other uncertainties in the spectral syntheses, in practice we have only adopted upper limits for the boron abundance in these two stars. This also provides an illustration of the limits to which we can trust our spectral synthesis. In this star, with our default line list, the blend just red-ward of the boron feature in ESL 16, (Fig. \ref{fig:first-uv}a), includes a bit too much absorption even in the absence of any boron. Reducing the  \ion{Mn}{3} 2066.546 \AA\ line by 0.35 dex and of the \ion{Ni}{3} 2066.824 \AA\ line by 0.3 dex (restoring the latter to its canonical value), resolves that discrepancy, (Fig.\ \ref{fig:first-uv}b), allowing a larger boron abundance. So for this star, and the similarly narrow-lined ESL 07, after consideration of such uncertainties, we adopt only upper limits to the boron abundance. 

\begin{figure}
    \gridline{
        \fig{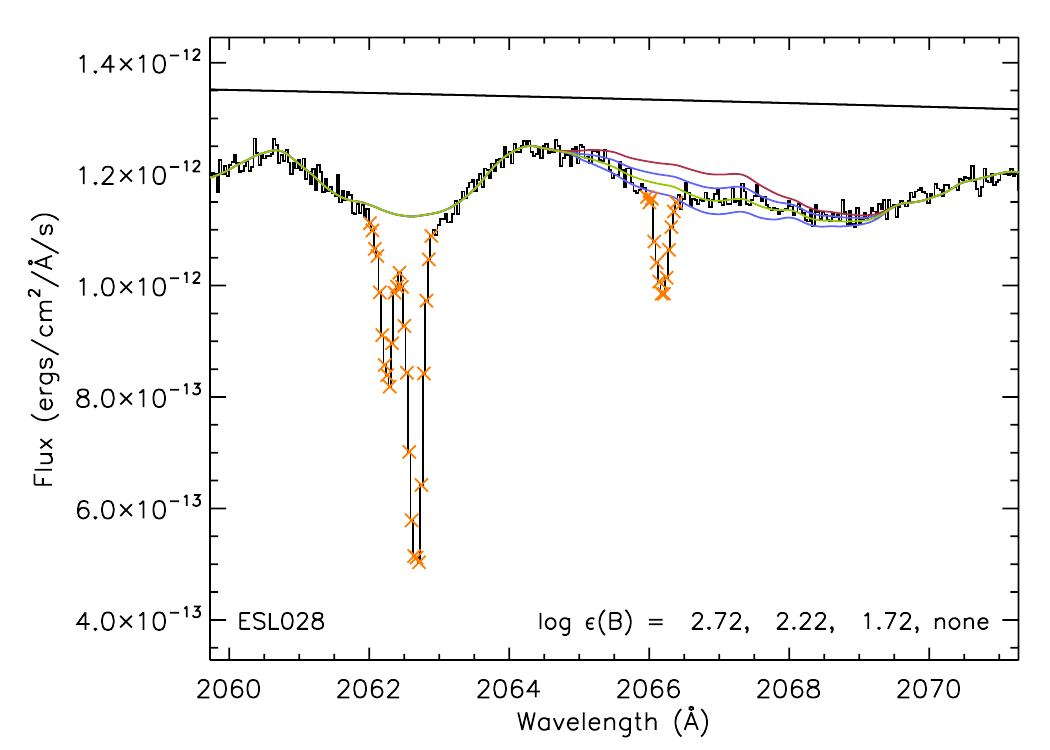}{0.49\textwidth}{(a) ESL 28}
        \fig{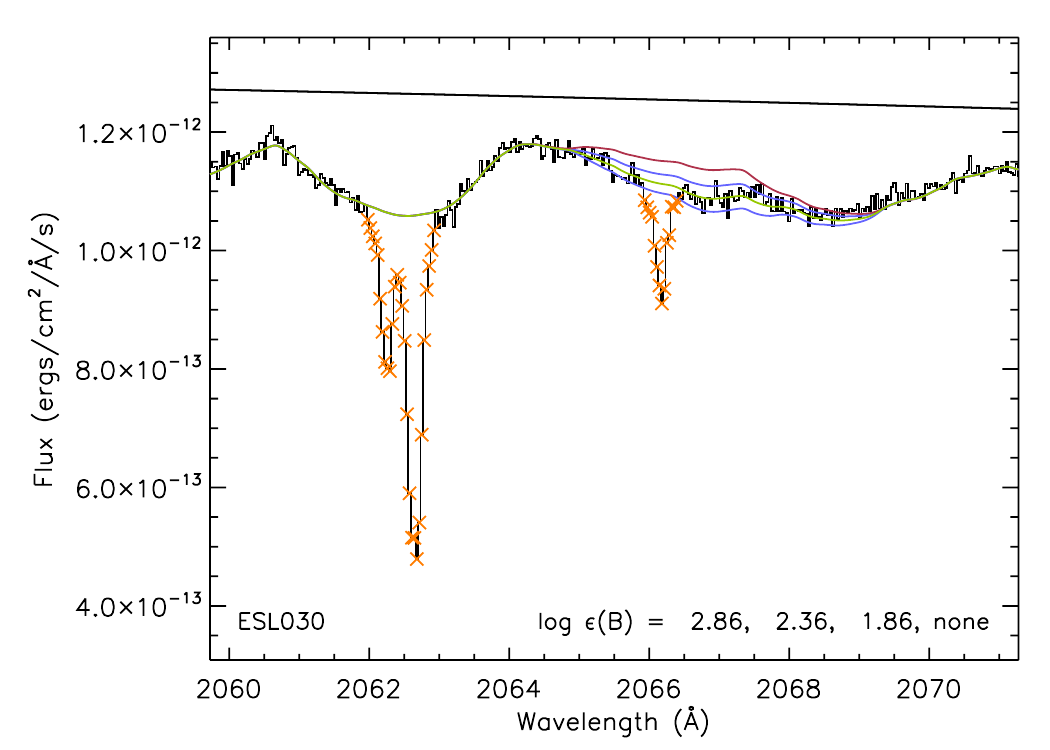}{0.49\textwidth}{(b) ESL 30}
    }
    \gridline{
        \fig{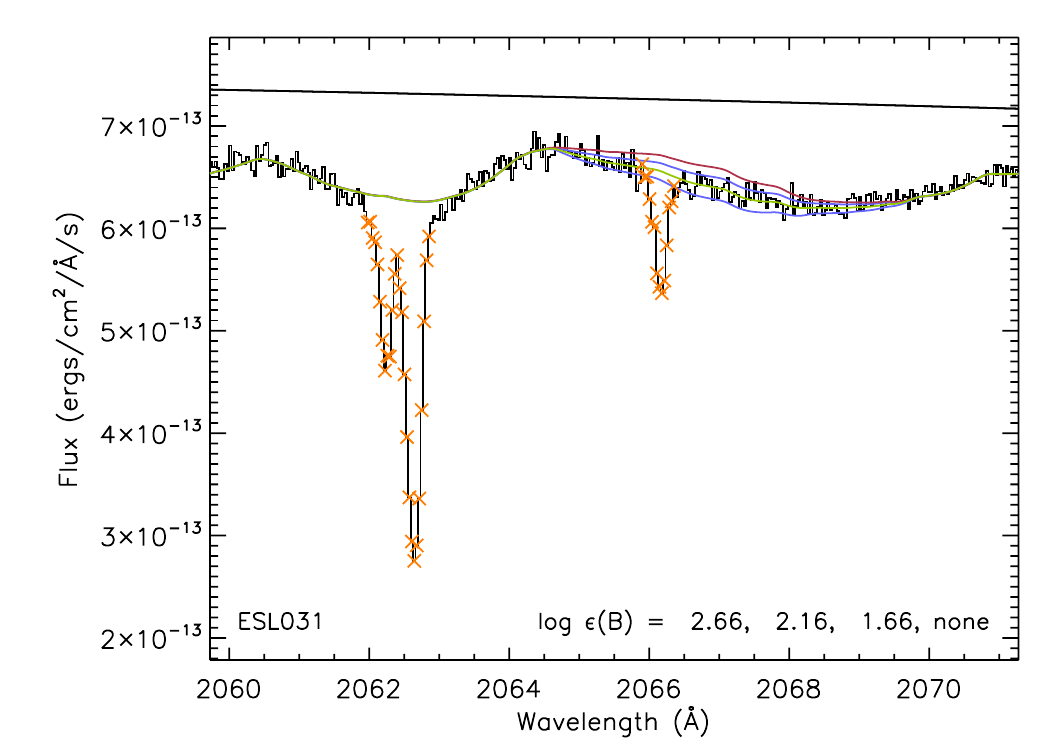}{0.49\textwidth}{(c) ESL 31}
        \fig{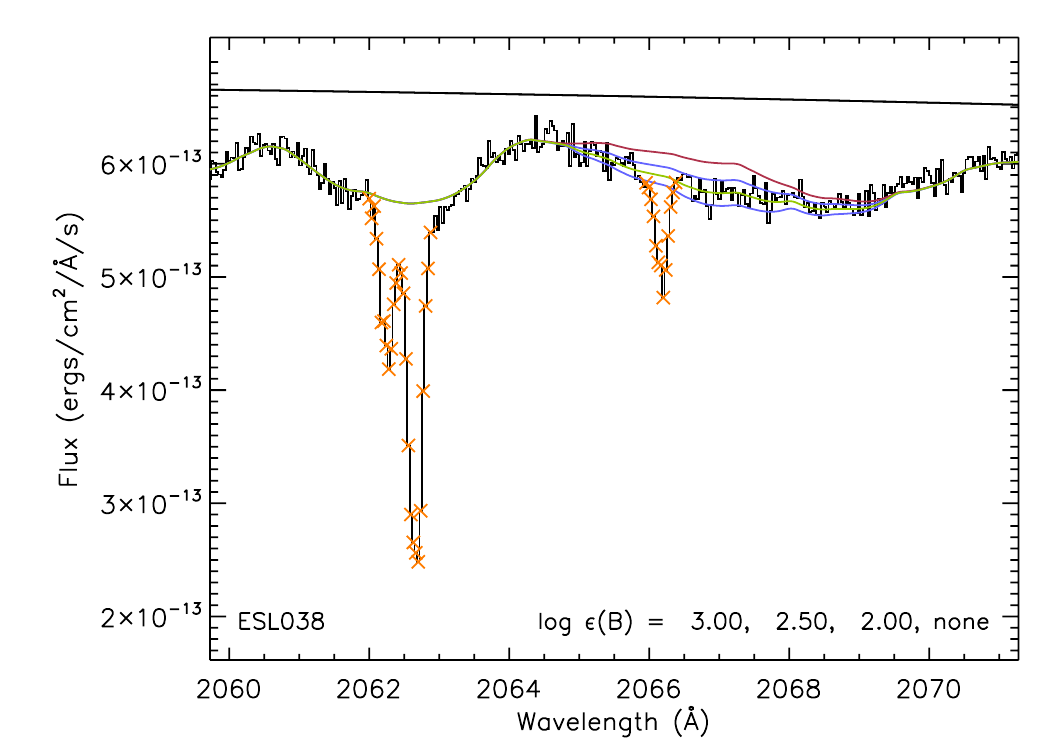}{0.49\textwidth}{(d) ESL 38}
    }
    \caption{The lowest mass stars in our sample. Line styles and other annotations are as in Fig.\ \ref{fig:first-uv}.}
    \label{fig:lowmass}
\end{figure}

For the lower mass stars in our sample, ESL 23, 24, 28, 30, 31, and 38, we find significantly larger boron abundances, and applying the same line list change we illustrated for ESL 16 would only increase the abundances by 0.1 to 0.15 dex.
Our fits for ESL 28, 30, 31, and 38 are shown in Figure \ref{fig:lowmass}, and our new fits to ESL 23 and 24 are qualitatively similar to those of \citet{2016ApJ...824....3P}. 

\begin{figure}
    \gridline{
        \fig{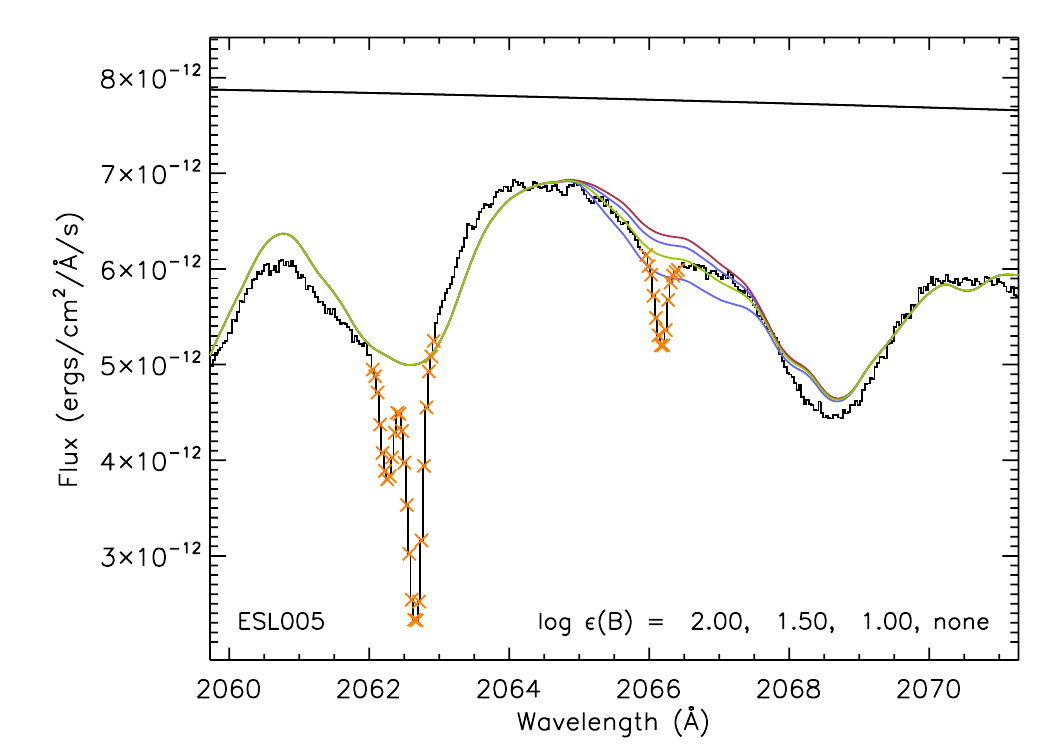}{0.49\textwidth}{(a) ESL 05}
        \fig{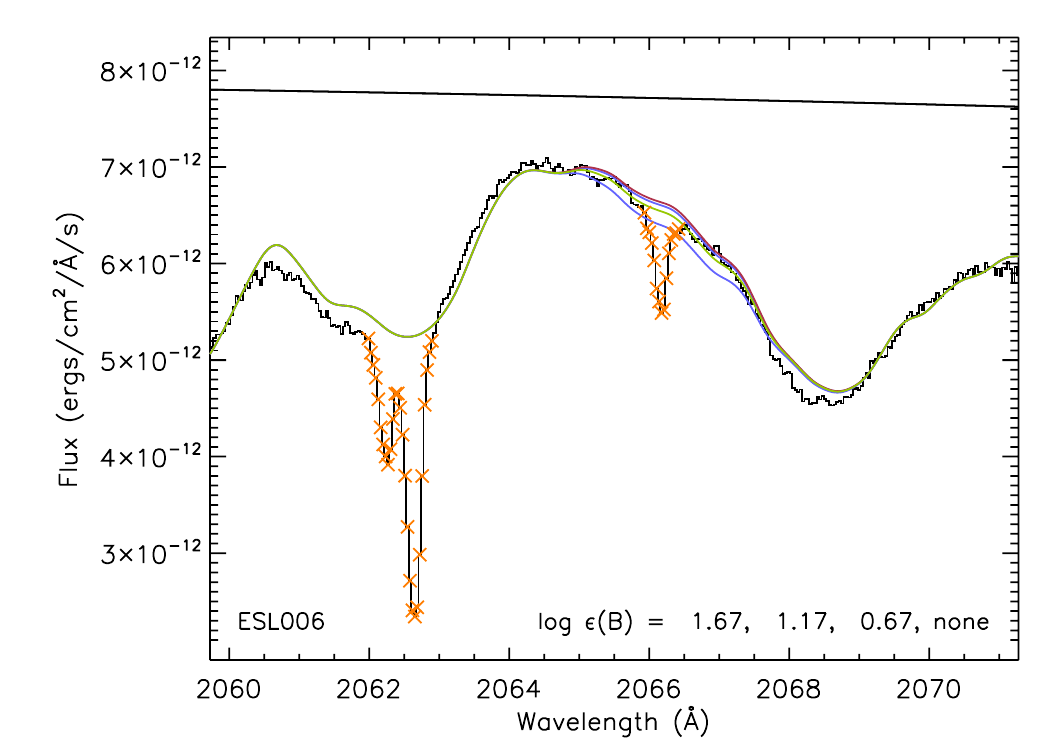}{0.49\textwidth}{(b) ESL 06}
    }
    \gridline{
        \fig{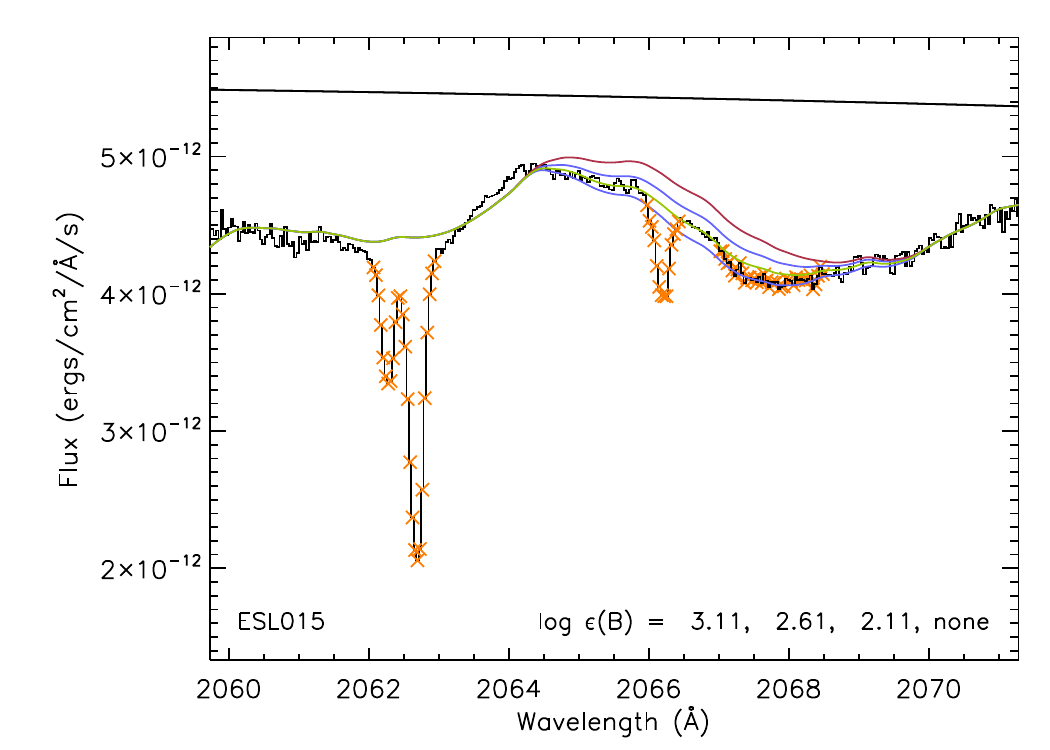}{0.49\textwidth}{(c) ESL 15}
        \fig{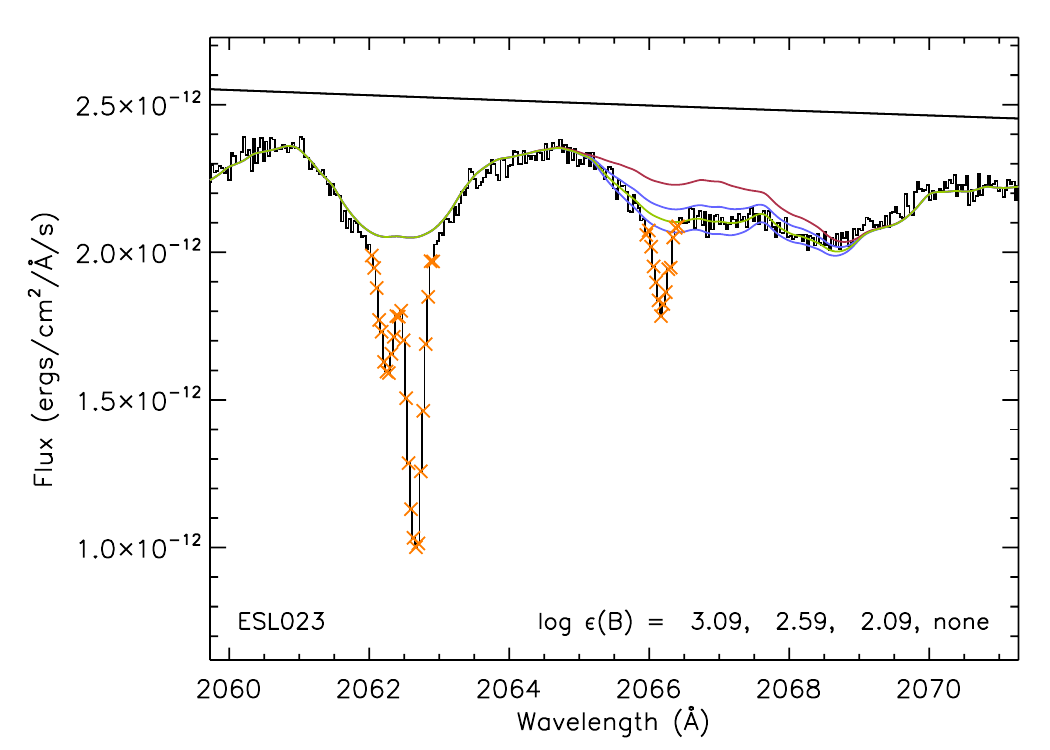}{0.49\textwidth}{(d) ESL 23}
    }
    \caption{Fits for ESL 5, 6, 15, and 23. Line styles and other annotations are as in Fig.\ \ref{fig:first-uv}, { except that for ESL 15 we also mark with orange crosses the extra region omitted from the final fit to the boron line}.}
    \label{fig:last-uv}
\end{figure}

The more evolved broad-lined stars are subject to additional uncertainties, due both to their apparently lower boron abundances and to the larger impact expected from rotational distortions. For ESL 05, (Fig. \ref{fig:last-uv}a), there does appear to be some additional opacity at the location of the boron line, but the overall shape of this spectral region is not well matched by the synthesis.  For ESL006, (Fig.\ \ref{fig:last-uv}b), the overall fit is somewhat better, but the best fit boron abundance can not be distinguished from zero. Both stars show extra absorption in the blue wing of the strong blend centered around the \ion{Ti}{4}\ 2066.241 \AA\ and \ion{Fe}{3} 2066.915\AA\ lines.  In ESL 15, (Fig. \ref{fig:last-uv}c), this extra absorption near 2067 \AA\ is even more prominent and we have excluded the affected region from our fit to the boron feature.

\cite{Hunter2009} had measured nitrogen abundances for most stars in their NGC 3293 sample, finding that, apart from ESL 01 and 02, there was otherwise no evidence for enhanced nitrogen in any of their stars. {  \citet{Morel2022} finds slightly higher mean abundances in NGC 3293 than did Hunter for both nitrogen (7.72 vs 7.60 respectively) and carbon (8.13 vs 7.97). We  list the carbon and nitrogen abundances they found for our individual program stars in Table \ref{tab:prevwork}.
While for the most part the two sets of results show similar trends, for a few individual stars \citet{Morel2022} finds nitrogen abundances that appear to be slightly enhanced. The departure is most significant for ESL 30, where Morel found an abundance of $8.01\pm 0.12$ vs Hunter's result of $7.57\pm 0.27$. However, Morel also finds a similarly enhanced carbon abundance for this star and so the carbon to nitrogen ratio is still close to the mean. Less significant apparent enhancements are also found by Morel for ESL 23 and ESL 24.  However, it should also be remembered that for all four of these stars, Morel adopted a fixed value of $\xi$ that is significantly smaller than the value determined by Hunter, or by us above in section \ref{sect:optanal}, and this will impact the derived abundance.}

\begin{figure}{ht}
    \centering
    \includegraphics[width=0.45\linewidth]{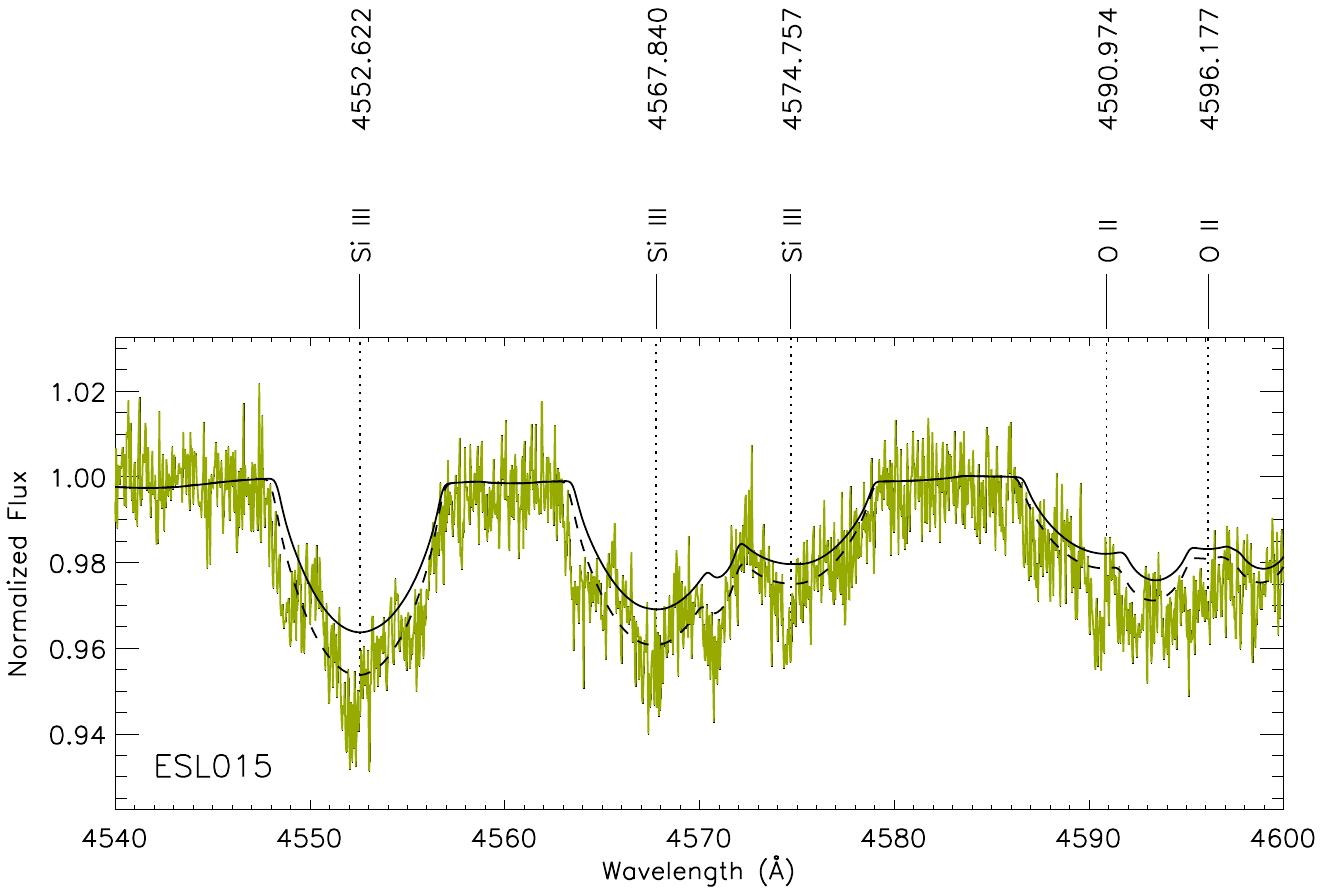}
    \caption{The observed FEROS spectrum of ESL 15 is compared to synthetic spectra around the \ion{Si}{3}\ triplet near 4560 \AA.  The solid line shows the synthetic spectrum calculated assuming the parameters given in Table \ref{tab:optical}, while the dashed lines shows the result of increasing the adopted micro-turbulence from 6 to 10 km/s to optimize the overall fit to these features.  }
    \label{fig:esl015_Si_III}

\end{figure}

However,  neither \citet{Hunter2009} nor\ \citet{Morel2022} derived a nitrogen  abundance for ESL 15, and in our initial attempt to do this, we  were surprised when a preliminary fit to the 3995 \AA\ \ion{N}{2} feature using the parameters in Table \ref{tab:optical} suggested a nitrogen enhancement of about 0.3 dex relative to the other stars, especially given our apparent detection of a large boron abundance for this star.  However, a closer examination of the optical spectrum suggests that many of the lines are affected by additional narrow absorption features that are not reproduced by the broad-lined synthetic spectrum.  This can be seen, for example, in the \ion{Si}{3} triplet near 4560 \AA\ (see \ref{fig:esl015_Si_III}), where there appear to be additional sharp features shifted with respect to the best fit synthetic spectrum by approximately $-240$, $-20$, and $+180$\,km/s, with the same features appearing to repeat in each member of the triplet.  Such line profile variations have been seen in other $\beta$~Cephei stars, (e.g.\ \citet{1998A&A...339..150T, 2004A&A...416.1069S}. Similar effects in the weaker and noisier \ion{N}{2} line could easily cause an overestimate of that line's equivalent width. Until the impact of such line profile variations on abundance determinations can be taken into account, results for this star in particular should be treated with caution. While a number of our other targets are $\beta$\ Cephei variables, none of these appear to show such obvious distortions of the line profiles.

{ Based on our current evaluation of the data, we conclude that there is no convincing evidence for significant nitrogen enhancement in any of our program stars apart from ESL 02 and possibly ESL 15. However, a more detailed reexamination of all of the available nitrogen lines to set tighter limits, including a more consistent understanding of micro-turbulence, would be worthwhile.}

\section{Comparison with Stellar Model Predictions}
\label{sec:comp}

The boron abundance measurements obtained and reported here show that boron is at least partially depleted in most of the early B-type main sequence stars in this cluster. While this is consistent with earlier measurements of boron in many similar field stars, below we will discuss the constraints these observations put on stellar evolution models thanks to their cluster membership.

Since rotationally induced mixing has been suggested as depletion mechanism, \citep{1996A&A...308L..13F, 1996A&A...307..849V, 2010A&A...522A..39F, 2010pese.confE..17L, 2011A&A...530A.115B}, we computed a dense grid of rotating single star models with MESA, \citep{Jin2023a}, with most of the physics assumptions as in the work of \citet{2011A&A...530A.115B}. { For our initial calculations, we used the canonical rotational mixing efficiency based on a theoretical work by \citet{1992A&A...253..173C} that is widely adopted in the literature, \citep[$f_c=0.033$, cf.\ ][]{2000ApJ...528..368H, 2017PhDT.......434M, 2021MNRAS.507.5013M, 2022MNRAS.517.2028K}. Note that \citet{2011A&A...530A.115B} adopted $f_c=0.0228$, a value slightly weaker than the canonical value.} We then used these models to produce synthetic populations for comparison with our observational data. These predictions are based on a population synthesis assuming a \citet{1955ApJ...121..161S} initial mass function, a distribution of initial rotational velocities for early B dwarfs in the Tarantula region of the LMC given by \citet{2013A&A...550A.109D}, and an initial boron abundance of 2.76, (matching the initial estimated solar abundance of \citet{2021A&A...653A.141A}).

The adopted initial rotational velocity distribution is bimodal, with about one quarter of the stars being slow rotators with a distribution peak at $v_{rot}$ $\simeq 60\,$km\,s$^{-1}$, and three quarters are fast rotators with a distribution peaking at $v_{rot}$ $\simeq 300\,$km\,s$^{-1}$, \citep[see tab.\,6 of ][]{2013A&A...550A.109D}. Similar bimodal rotational velocity distributions are also found in young star clusters, where they manifest as split main sequences in high precision color-magnitude diagrams, \citep[e.g.,][]{2023A&A...672A.161M, 2023A&A...670A..43W},  and where they are also spectroscopically verified, \citep{2023MNRAS.518.1505K, 2023arXiv230814799B}.
Notably, the origin of the bimodality is still under debate, 
\citep[cf.\ ][]{2020MNRAS.495.1978B, 2022NatAs...6..480W}, and the boron surface abundance distribution may hold interesting clues.

\begin{figure*}
 \centering
\includegraphics[width=0.49\linewidth]{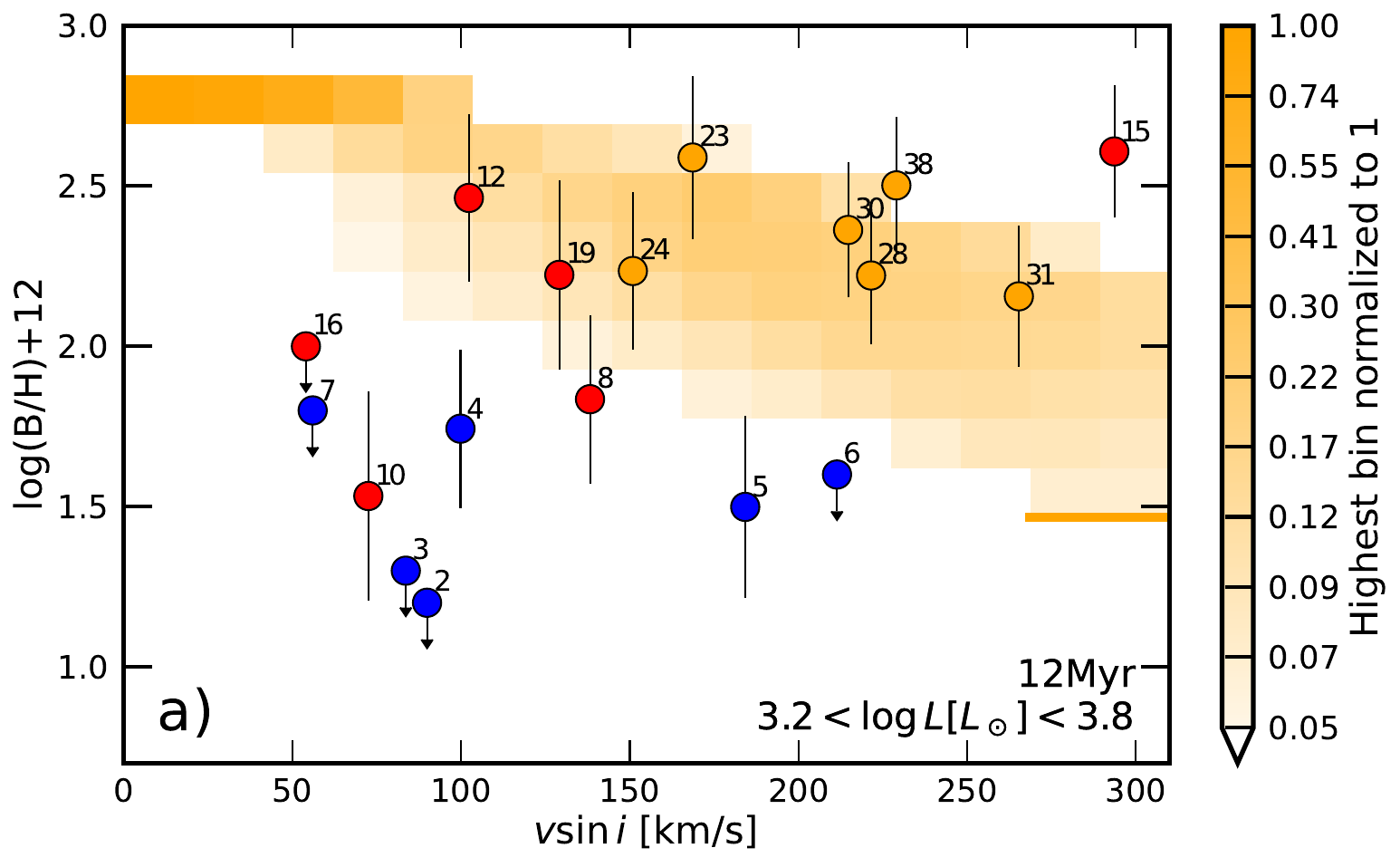}
\includegraphics[width=0.49\linewidth]{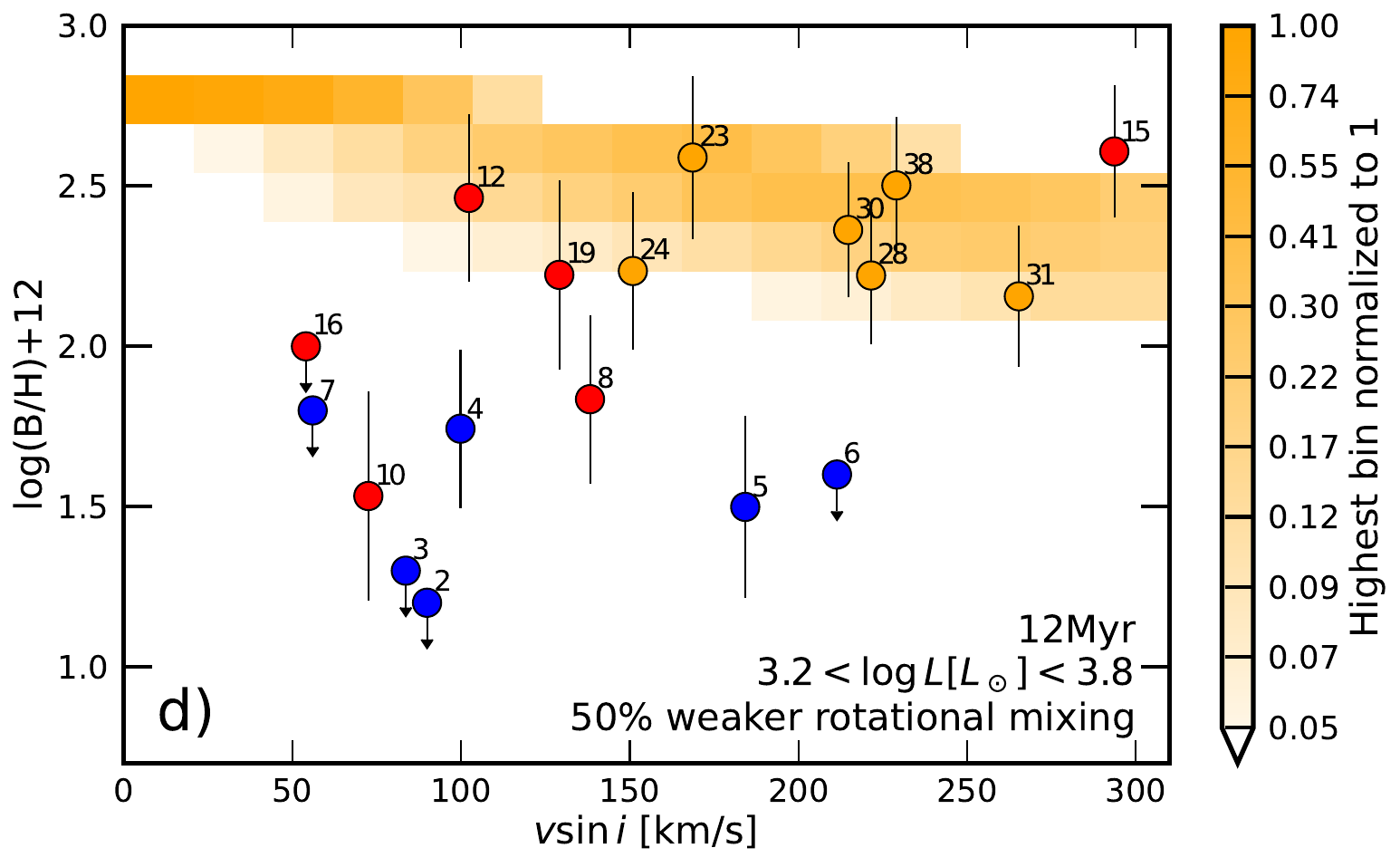}
\centering
\includegraphics[width=0.49\linewidth]{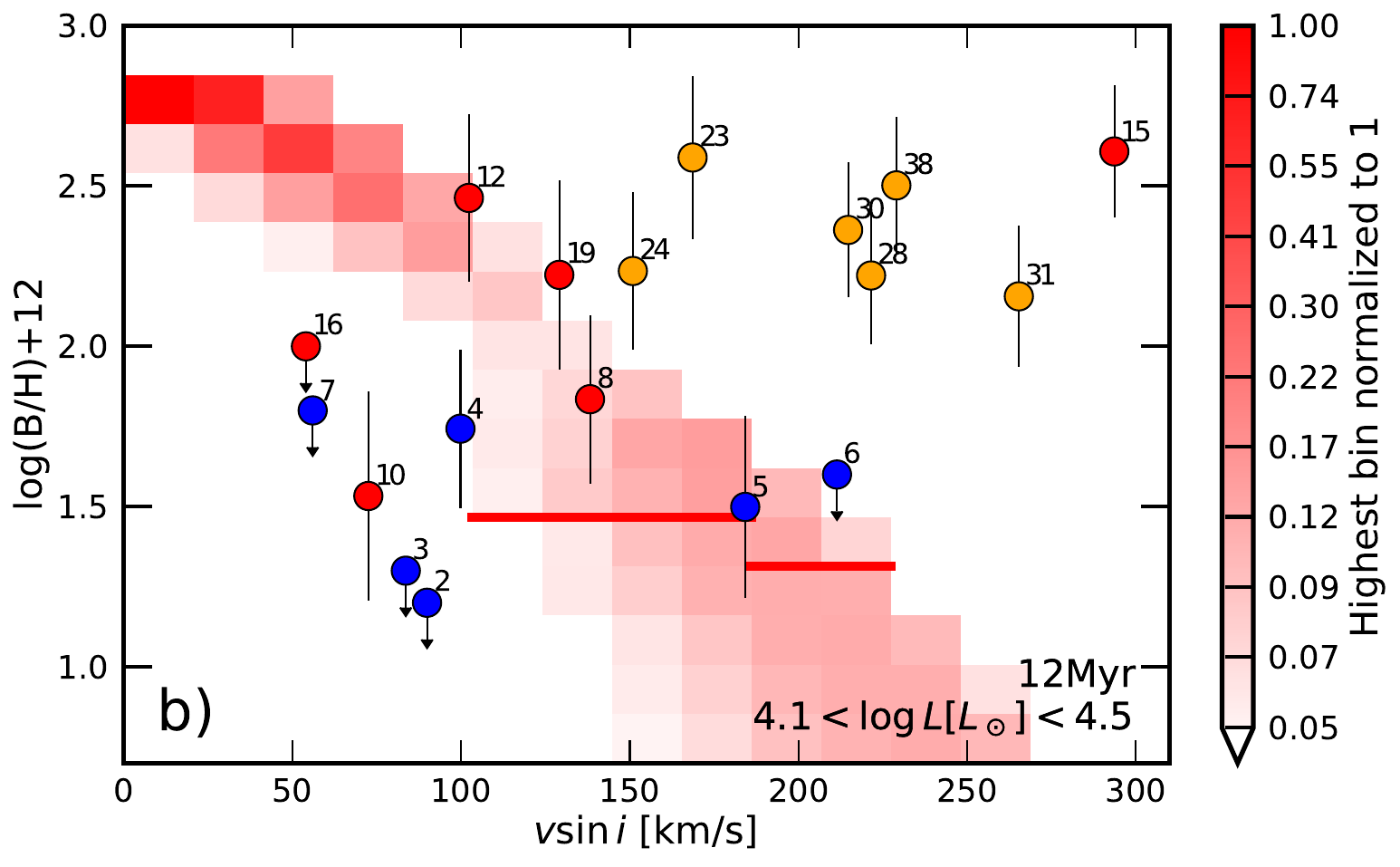}
\includegraphics[width=0.49\linewidth]{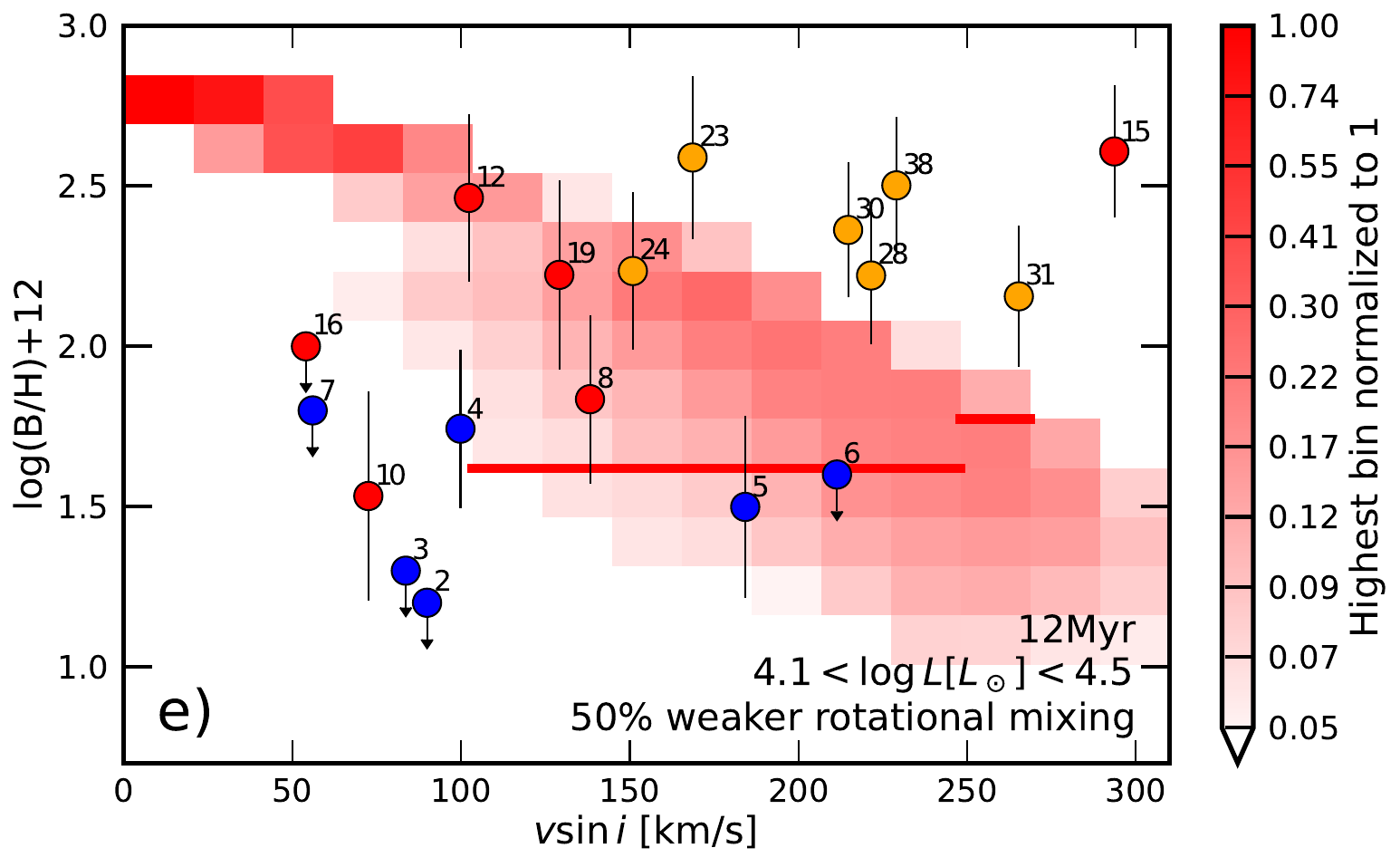}
\centering
\includegraphics[width=0.49\linewidth]{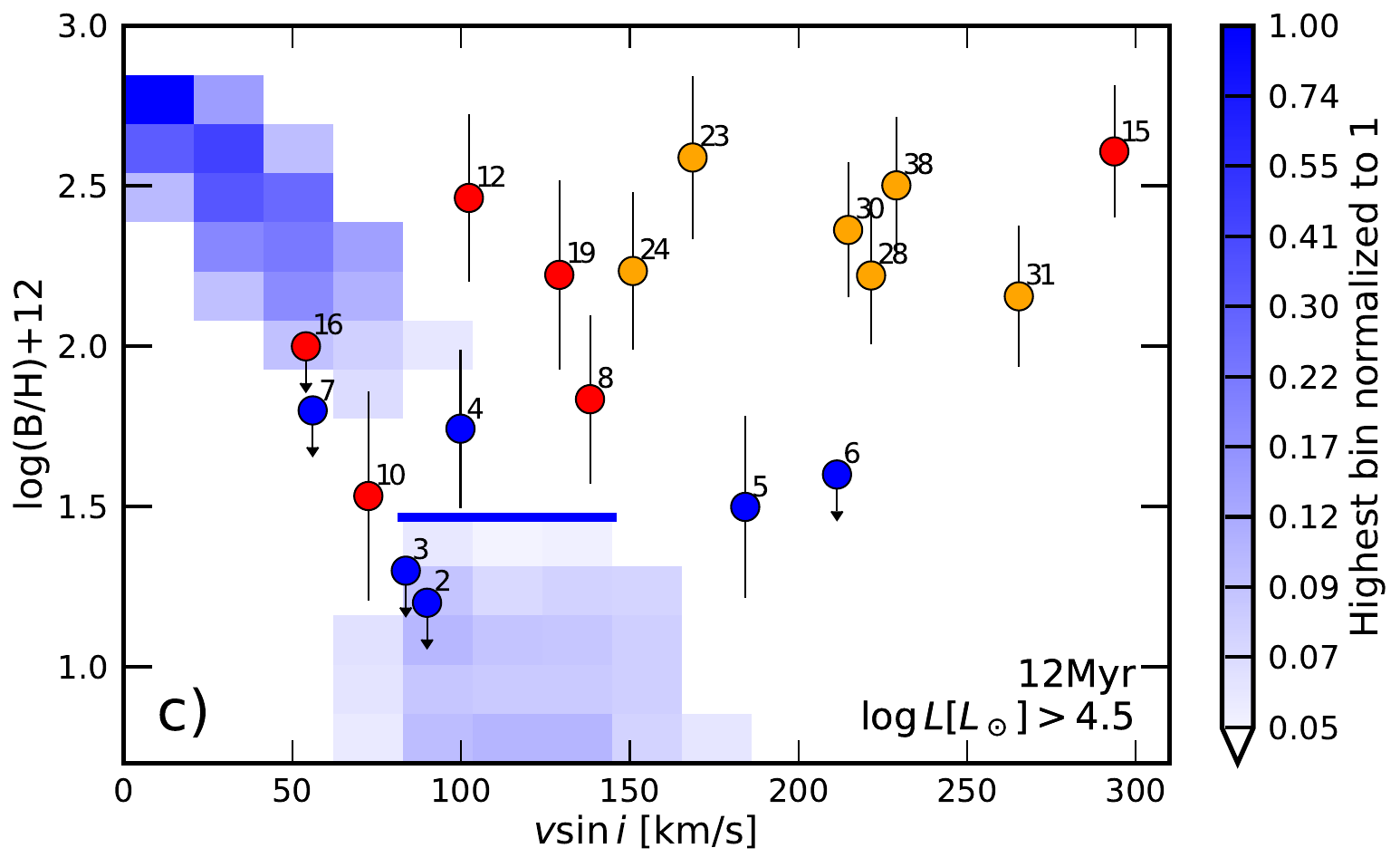}
\includegraphics[width=0.49\linewidth]{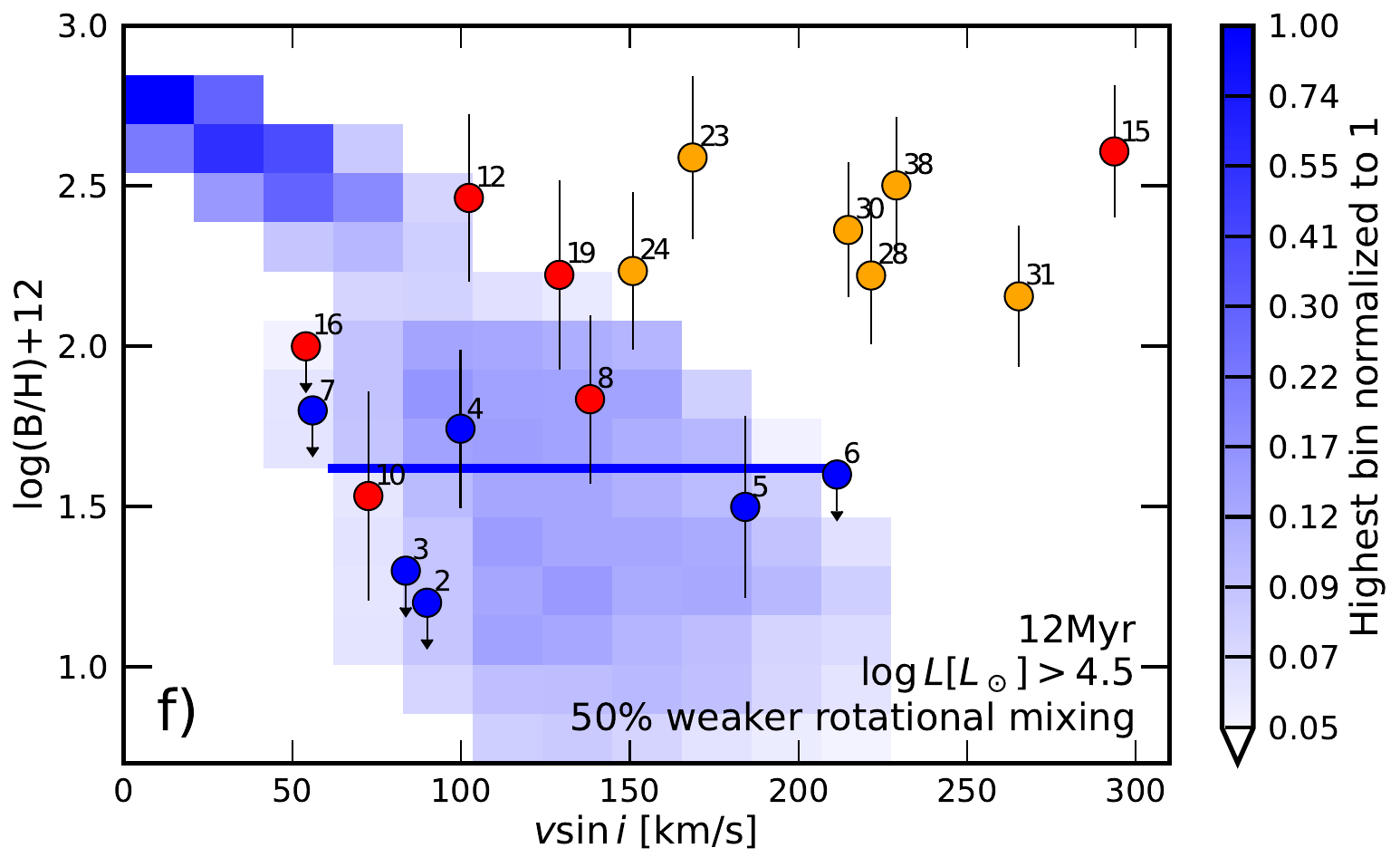}
    \caption{Boron abundances, or upper limits, for the investigated sample
    of stars in the open cluster NGC 3293 (colored circles), are plotted as a function of their projected rotational velocity. The colored background pixels represent the number distribution of a coeval 12 Myr old, single star population of the luminosity range indicated in the legend, with different luminosity
    ranges shown in each of the three panels. The color of a circle indicates to which of the three luminosity intervals the corresponding star belongs. In each synthetic population,
    in the pixel immediately above the colored horizontal line, less than 10\% of the stars are nitrogen-enriched
    by more than 0.2\,dex. 
    { Panels\,a,\,b,\,and c show the predicted synthetic population calculated using the standard mixing efficiency coefficient, $f_c=0.033$, while panels\,d,\,e,\,and f show the results of reducing the mixing efficiency by 50\%, $f_c=0.0165$, when constructing the synthetic population.}}
   
    \label{fig:b_vs_rot}
    \end{figure*}

To test the concept of boron depletion by rotational mixing, a Hunter-type diagram (surface abundance versus $v \sin i$) appears well suited. It has been employed repeatedly using the nitrogen surface abundance,
\citep{2008A&A...479..541H, Hunter2009, 2011A&A...530A.115B, 2015A&A...577A..23M, 2017A&A...600A..82G}, as well as boron, \citep{2010A&A...522A..39F, 2010pese.confE..17L}.
We show in Fig. \ref{fig:b_vs_rot} the boron Hunter diagram of all main sequence stars in NGC 3293 for which boron abundances have been measured or constrained so far. Since faster rotation is thought to lead to faster boron depletion, a corresponding trend may be expected, but the bulk of the data does not show such a trend.
However, a trend of a decreasing boron abundance for stars with larger $V \sin i$ is expected for an unbiased massive single star population produced via constant star formation, even though with considerable scatter, \citep{2010pese.confE..17L}. Our sample deviates from such a population. Instead of sampling random ages, our targets belong to an open star cluster, and can be therefore considered as coeval. As such, they contain stars near the cluster turn-off, which are close to ending core hydrogen burning, and stars of roughly half the turn-off mass and bolometrically 30-times fainter (cf., Fig.2), which so far consumed just less than 10\% of their initial hydrogen in their core.
Stellar models including rotational mixing do predict a faster boron depletion in more rapidly rotating stars. 
In addition, the depletion timescale drops rapidly with stellar mass, such that in two coeval stellar models with the same rotational velocity, the more massive one is expected to show more boron depletion, \citep[cf., fig.\ 11 of ][]{2011A&A...530A.115B}.

To illustrate this, we split the boron-Hunter diagram into three panels (Fig.\,\ref{fig:b_vs_rot}, Panels a, b, c), corresponding to three luminosity ranges which contain the investigated stars (cf., Fig.\,\ref{fig:hrd}). The synthetic sub-populations in these three luminosity intervals form rather linear streaks in the Hunter plots, which are all anchored at the solar boron abundance for $v \sin i=0$, and which become steeper at higher luminosities. These streaks sweep almost through the whole area of the Hunter plot, and reflect the strong dependence of the surface boron abundance in coeval rotating main sequence star models on mass or luminosity. It can be seen that the corresponding observed sub-populations roughly follow the trends suggested by the synthetic sub-populations (except for ESL 15, which we ignore here for reasons pointed out in Sect.\,4). Our data may therefore indeed be suited to validate and constrain rotational mixing, as we pursue below.

A closer look shows, however, that the slowest rotators in our sample are all strongly boron depleted. In particular for the mid- and high-luminosity groups, our models can not reproduce all observed stars with the
same mixing efficiency and age. { While ESL\,2 and\,3 are marginally reproduced by our synthetic population, the nitrogen enhancement in ESL\,2, \citep{Hunter2009}, implies that its boron abundance might be well below the determined upper limit.} 
We consider this as an indication that  a different boron depletion mechanism operated in the slow rotators, and postulate here, that the five stars with the smallest v sin i (ESL 16, 7, 10, 3 and 2) are intrinsic slow rotators, i.e., belong to the 25\% of stars with a peculiar formation history, which may not be reflected by our single star models. While for each individual star this can not be proven, the likelihood that, e.g., ESL 16 belongs to the population of rapid rotators is less than 20\%.

The remaining stars do show both expected trends, a stronger boron depletion for faster rotation and for higher luminosity, as predicted by our models, which can be exploited quantitatively. A close inspection of panels a-c of Fig.\,\ref{fig:b_vs_rot} shows that in all three luminosity ranges, less mixing than adopted in the stellar models would lead to a better agreement. This is even more so when an older age of NGC 3293 is considered. 
{ Panels\,d, e, and f  in Fig.\,\ref{fig:b_vs_rot} illustrate the effect} of reducing the efficiency parameter for rotational mixing by 50\%, ($f_c = 0.0165$). Notably, an older age can be compensated by a smaller mixing efficiency. Since an age of 12 Myr is at the lower end of the most recently proposed ages for NGC3293, \citep[cf., ][]{Morel2022}, we conclude that rotational mixing is likely less efficient than assumed in our models. Assuming we do interpret the small additional dip in the spectrum of ESL 05 (Fig.\ \ref{fig:last-uv}a, relative to the otherwise similar ESL 06 (Fig.~\ref{fig:last-uv}b) as a detection of boron, then the resulting boron abundance of this star may provide the strongest constraint on a reduced rotational mixing efficiency.

The above analysis shows that stellar models including rotational mixing can reproduce both observed trends of the boron abundance with rotation and mass or luminosity. This supports the idea of rotational mixing being the dominating boron depletion mechanism in the rapidly rotating majority of early B type stars. Assuming a different boron depletion mechanism for the slow rotators helps this view, and sheds new light on their origin.

This population was uncovered by \citet{2008ApJ...676L..29H}, who were the first to identify a group of intrinsically slowly rotating early B dwarfs with a strong diversity in nitrogen enrichments, such that the nitrogen enrichment could not be associated to rotational mixing. While boron is not measured in these stars, it can be argued that any nitrogen enrichment process is likely to lead to boron depletion, since the CNO cycle operates at temperatures at which boron is completely destroyed. The boron measurements in the slow rotators of our sample provide new, strong constraints on the evolutionary history of such stars, because in some of them, boron is only partly depleted (ESL 10, and perhaps ESL 04), and nitrogen is not enhanced \citep[see also][]{2008A&A...481..453M}.

Perhaps, ESL 2 in our sample, and to a lesser extent also ESL 3, and 7, { are the most interesting} in this context. ESL 2 is an apparently slow rotator, strongly boron depleted, and the only star in our sample which is { clearly} nitrogen-enriched. Its location in the HR diagram clearly identifies it as a blue straggler (Fig. 2). Blue stragglers are thought to be the products of binary mass transfer or stellar mergers, 
\citep{2015ApJ...805...20S}. According to recent models, merger products in the mass range considered here may be slow rotators \citep{2019Natur.574..211S} and show various degrees of nitrogen enrichments, from none to strong, \citep{2013MNRAS.434.3497G}. 
\citet{2022NatAs...6..480W} have also proposed that the slow rotators which form the bluer main sequence in young star clusters originate from stellar mergers.
The boron abundances found in our sample stars may therefore fit into a consistent picture, with the majority of fast rotators enriching nitrogen and depleting boron by rotational mixing, and the slow rotators having a history of strong binary interaction. Slow rotators in which boron is not fully destroyed could support the suggestion of pre-main sequence mergers,
\citep{2020MNRAS.491.5158T, 2022NatAs...6..480W}. To turn this idea from a working hypothesis into a theory, significantly more boron measurements, as well as merger models that include proton captures on boron, are required.

One possible caveat to the arguments above, is that, as discussed by \citet{2000ApJ...528..368H}, rotational mixing can result from several distinct processes, and so adjusting the rotational mixing efficiency by the tuning of a single overall parameter may lead to misleading results when comparing stars in different regimes where the relative contributions of the various rotationally driven instabilities may differ. Exploration of this possibility will require additional theoretical work and will not be further discussed here.

\section{Discussion and conclusions}

The measurements of boron and nitrogen surface abundances in main sequence B\,type stars of this young open star cluster provide a unique opportunity to shed light on the origin of contamination of a significant fraction of these stars with the products of hydrogen fusion. The best quantitative interpretation of these measurements in terms of rotationally induced mixing in the stellar interior requires us to disregard the five most slowly rotating of our targets, which we tentatively consider as binary products where mixing mechanisms other than rotation dominate. We can then conclude that rotational mixing appears likely as explanation of the observed abundances of the remaining stars, since the trends of the surface abundances with stellar rotation and luminosity are in good agreement with the observations. 
{ We find the boron data is better reproduced with a rotational mixing efficiency parameter which is reduced by a factor of two ($f_c \approx 0.0165$), with respect to the widely adopted value of $f_c \approx 0.033$.}

The binary interpretation of the slow rotators is only based on circumstantial evidence (cf.\, Sect.\ref{sec:comp}). A rigorous analysis including the surface abundances of the slow rotators must await corresponding models predictions (Jin et al., in prep.). Nevertheless, we were able to construct a working hypothesis that the majority of rapidly rotating upper main sequence stars undergo mild rotationally induced mixing while the slow rotators are mostly merger products. While this is free of immediate contradictions, it is based on only 18 total stars, and so our conclusions may be sensitive to statistical fluctuations. It also leaves out the fastest rotators in young open clusters, i.e., the Be stars, which have been suggested to originate from mass transfer in binaries 
\citep{1991A&A...241..419P, 2021A&A...653A.144H}, and the roles of { gravity waves} \citep{2023ApJ...942...53V} and magnetic fields \citep{2011A&A...525L..11M}, which could amplify or reduce internal mixing. Further observations and stellar interior calculations are thus needed to come to more definite conclusions.

Acknowledgements
Based on observations with the NASA/ESA Hubble Space Telescope obtained [from the Data Archive] at the Space Telescope Science Institute, which is operated by the Association of Universities for Research in Astronomy, Incorporated, under NASA contract NAS5-26555. Support for program number 14673 was provided through a grant from the STScI under NASA contract NAS5-26555.
Much of the data presented in this paper were obtained from the Mikulski Archive for Space Telescopes (MAST) at the Space Telescope Science Institute. The specific observations analyzed can be accessed via \dataset[https://doi.org/10.17909/7vgt-zq51]{https://doi.org/10.17909/7vgt-zq51}. STScI is operated by the Association of Universities for Research in Astronomy, Inc., under NASA contract NAS5–26555. Support to MAST for these data is provided by the NASA Office of Space Science via grant NAG5–7584 and by other grants and contracts. SD acknowledges CNPq/MCTI for grant 306859/2022-0. 
DJL is supported by the Spanish Government Ministerio de Ciencia e Innovaci\'on through grants PID2021-122397NB-C21 and SEV 2015-0548.
HJ received financial support for this research from the International Max Planck Research School (IMPRS) for Astronomy and Astrophysics at the Universities of Bonn and Cologne.
The authors gratefully acknowledge the granted access to the Bonna cluster hosted by the University of Bonn.
This paper also made use of data obtained from the ESO Science Archive Facility and observed using ESO Telescopes at the La Silla Paranal Observatory under ESO observing program IDs 73.D-0234, 171.D-0237, 081.A-9006,  081.A-9008, and 188.B-3002.
We want to thank the anonymous referee for providing a careful review and insightful comments that resulted in significant improvements to the final paper.
We also thank Jes\'us Ma\'iz Apell\'aniz for useful discussions on the distance to NGC\,3293.

\bibliography{ngc3293-2024-accepted}{}
\bibliographystyle{aasjournal}



\end{document}